\documentclass[11pt]{article}
\usepackage{jheppub}
\usepackage{amsmath,amssymb,amsfonts,graphicx,slashed,amsthm,mathtools,upgreek, enumerate, tensor, subfig}
\usepackage[dvipsnames]{xcolor}
\usepackage{arydshln}

\usepackage{comment}
\usepackage{hyperref}
\usepackage[utf8]{inputenc}
\usepackage[titletoc]{appendix}

\numberwithin{equation}{section} 
\allowdisplaybreaks

\allowdisplaybreaks

\newcommand{\AK}[1]{{\color{red}{{\bf AK:} #1}}}

\newcommand{\be}{\begin{equation}}
\newcommand{\ee}{\end{equation}}
\newcommand{\f}{\frac}
\newcommand{\s}{\sqrt}
\newcommand{\p}{\partial}

\newcommand{\bea}{\begin{eqnarray}}
\newcommand{\eea}{\end{eqnarray}}
\newcommand{\ba}{\begin{align}}
\newcommand{\ea}{\end{align}}

\newcommand{\la}{\langle}
\newcommand{\ra}{\rangle}
\newcommand{\beq}{\begin{equation}}
\newcommand{\eeq}{\end{equation}}
\newcommand{\bra}[1]{\langle #1 |}
\newcommand{\ket}[1]{| #1 \rangle}
\newcommand{\inner}[2]{\langle #1 | #2 \rangle}
\newcommand{\avg}[1]{\langle #1 \rangle}

\DeclareMathOperator{\tr}{tr}

\title{
 Entanglement between  two   gravitating universes
}

\author[a, b]{Vijay Balasubramanian}
\author[c]{\!, Arjun Kar}
\author[d,e]{\!, Tomonori Ugajin}

\affiliation[\,a]{David Rittenhouse Laboratory, University of Pennsylvania,\\
209 S.33rd Street, Philadelphia, PA 19104, USA}
\affiliation[\,b]{Theoretische Natuurkunde, Vrije Universiteit Brussel (VUB), and \\ International Solvay Institutes, Pleinlaan 2, B-1050 Brussels, Belgium}
\affiliation[\,c]{Department of Physics and Astronomy, University of British Columbia,\\
6224 Agricultural Road, Vancouver, BC V6T 1Z1, Canada}
\affiliation[\,d]{Center for Gravitational Physics,
Yukawa Institute for Theoretical Physics, Kyoto University,\\
Kitashirakawa Oiwakecho, Sakyo-ku,
Kyoto 606-8502, Japan}
\affiliation[\,e]{The Hakubi Center for Advanced Research, Kyoto University,\\
Yoshida Ushinomiyacho, Sakyo-ku, Kyoto 606-8501, Japan}
\emailAdd{vijay@physics.upenn.edu}
\emailAdd{arjunkar@phas.ubc.ca}
\emailAdd{tomonori.ugajin@yukawa.kyoto-u.ac.jp}

\preprint{YITP-21-39}

\abstract{We study two disjoint  universes in an entangled pure state.  When only one universe contains gravity, the path integral for the $n^{\text{th}}$ R\'enyi entropy includes a  wormhole between the $n$ copies of the gravitating universe, leading to a standard ``island formula'' for entanglement entropy consistent with unitarity of quantum information.
When both universes contain gravity, gravitational corrections to this configuration lead to a violation of unitarity.  
However, the path integral is now dominated by a novel wormhole with $2n$ boundaries connecting replica copies of both universes.
The analytic continuation of this contribution involves a quotient by $\mathbb{Z}_n$ replica symmetry, giving a cylinder connecting the two universes.
When entanglement is large, this configuration has an effective description as a ``swap wormhole'', a geometry in which the boundaries of the two universes are glued together by a ``swaperator''.  This description allows precise computation of a generalized entropy-like formula for entanglement entropy.
The quantum extremal surface computing the entropy lives on the Lorentzian continuation of the cylinder/swap wormhole, which has a connected Cauchy slice stretching between the  universes -- a realization of the ER=EPR idea. The new wormhole restores unitarity of quantum information.
} 


\begin{document}

\maketitle

\section{Introduction}

Recently, the connection between quantum entanglement  and the geometric structure of spacetime  has been studied extensively  \cite{Ryu:2006bv,Ryu:2006ef,Hubeny:2007xt,VanRaamsdonk:2010pw,Maldacena:2013xja}. A concrete realization of 
this idea is the ``island formula'' \cite{Almheiri:2019hni, Almheiri:2019psf,Penington:2019npb,Penington:2019kki,Almheiri:2019qdq}, which modifies the usual rule for computing entanglement entropy in the presence of gravity, enabling us to reproduce the expected Page curve for the Hartle-Hawking  state on an evaporating black hole  \cite{Page:1993wv,Page:2013dx}.  

The conventional setting for studying the island formula is  a black hole  in anti de Sitter space (AdS) attached to a non-gravitating heat bath at the spacetime boundary.  We impose transparent boundary conditions so that Hawking quanta emitted by the black hole  can escape to the bath.  Since gravity is turned off in the bath, we can define the entanglement entropy of a region $A$  within it in the usual manner as a function of the reduced density matrix on $A$.  The island formula says that the entanglement entropy of A involves a distinguished region  $I$ in the gravitating part of the spacetime \cite{Almheiri:2019hni},  
\be
S(\rho_{A}) = \underset{I}{\text{min ext}} \left[ \f{{\rm Area} (\p I)}{4G_{N}}  + S_\text{eff} (A\cup I ) \right] \, , \label{eq:islandformula}
\ee
where ${\rm Area}(\p I)$ denotes the  area of the boundary of $I$,  $S_\text{eff} (A\cup I)$ is the QFT entanglement  entropy of the effective bulk field theory on $A \cup I$, and we are to minimize and extremize over all such regions $I$. The region which extremizes the above entropy functional is called island.  The island formula is proved by using  the replica trick for  entanglement entropy  $ S(\rho_{A}) = \p_{n}  \tr \rho_{A}^{n}|_{n=1}$. The R\'enyi entropy  $\tr  \rho_{A}^{n}$ is computed by a gravitational path integral on $n$ copies of the original system, and has a novel saddle in the semiclassical  $G_{N} \rightarrow 0$ limit -- a wormhole connecting the $n$ replicas. This replica wormhole \cite{Penington:2019kki,Almheiri:2019qdq} leads to the island  formula in the in the $n \rightarrow 1$ limit. The island formula also has been  applied to  black holes in flat space  \cite{Hashimoto:2020cas,Hartman:2020swn, Anegawa:2020ezn, Krishnan:2020oun, Gautason:2020tmk,Wang:2021woy, Matsuo:2020ypv, Wang:2021mqq,Miyata:2021ncm} and de Sitter space \cite{Chen:2020tes,Hartman:2020khs,Balasubramanian:2020xqf,Sybesma:2020fxg,Aalsma:2021bit,Geng:2021wcq}.

It is challenging to extend this analysis directly to a situation where the heat bath is itself gravitating because  there is no diffeomorphism  invariant notion of a ``region" in general in the presence of gravity, and because the Hilbert space of quantum gravity cannot be factorized into a Hilbert space of a region and that of the complement, due to the edge modes living in the boundary of the region \cite{Donnelly:2011hn, Casini:2013rba,Donnelly:2016auv}.  That said,  interesting progress has been made in \cite{Dong:2020uxp,Geng:2020fxl,Geng:2021wcq, Anderson:2021vof,Geng:2021iyq}.
An alternative approach, following \cite{Balasubramanian:2020coy,Balasubramanian:2020xqf,Miyata:2021ncm}, is to start with two disjoint universes, $A$ and $B$, which support quantum field theories  $QFT_{A}$ and $QFT_{B}$, taken to be identical for simplicity.  Since A and B are disjoint, the Hilbert space of the total system is a tensor product $\mathcal{H}_{A} \otimes \mathcal{H}_{B}$, so one can define their entanglement as usual.   If only universe $B$ is gravitating, \cite{Balasubramanian:2020coy,Balasubramanian:2020xqf,Miyata:2021ncm} argued that  the entanglement entropy $S(\rho_{A})$ of the non-gravitating universe $A$ is again computed by an island formula similar to \eqref{eq:islandformula}.  In this case, the path integral for the $n^{{\rm th}}$ R\'{e}nyi entropy is dominated by a replica wormhole which connects the $n$ copies of the gravitating universe. Similar setups were also studied in \cite{Penington:2019kki,Hartman:2020khs}.

In this paper, we derive a formula for entanglement entropy by using the replica trick when both universes $A$ and $B$ are gravitating. There is no ambiguity in defining entanglement entropy even if gravity is present in both universes, because they are disjoint and hence we can avoid the subtleties of defining diffeomorphism invariant notions of a ``region'' and lack of factorization of Hilbert spaces.   If the cosmological constant is negative, we can be a bit more precise and say that the dual description of the gravitating theory has two distinct Hilbert space factors.  The gravitational theory we have mind is the effective low energy description that emerges of this UV-complete system.
In this setting, various new wormholes appear as saddles of  the gravitational path integral for the R\'enyi entropy. We show that when the entanglement is large, the path integral for the $n^{{\rm th}}$ R\'{e}nyi entropy is  dominated by a wormhole that connects the replica copies of  both universes -- i.e. it has a total of $2n$ asymptotic regions.

When both $A$ and $B$ are gravitating, the contribution of the fully connected wormhole $M_{2n}$  has a complicated form, and  its von Neumann entropy limit $n \rightarrow 1$  cannot always be interpreted as a generalized entropy, i.e. a formula like the right hand side of \eqref{eq:islandformula}.
However,  when the entanglement between $A$ and $B$ is sufficiently large, there is a simplification. We show that, in this limit, the entanglement entropy $S(\rho_{A})$ is again computed by  a generalized entropy \eqref{eq:islandformula} on a spacetime.  However, the relevant  geometry is neither the original universe $A$ nor $B$.  Rather it is a {\it new} spacetime $A/B$ constructed by gluing $A$ and $B$ together.  This is reminiscent of the ER=EPR idea \cite{Maldacena:2013xja}, which relates quantum entanglement to  wormholes in spacetime.  In the Euclidean picture, the effective wormhole, which we call a ``swap wormhole'', attaches both the bulk and the boundary of universe $A$ to the bulk and boundary of universe $B$. We insert an effective boundary condition-changing operator, or ``swaperator", at the location on the spacetime boundary of this wormhole where we transition from the boundary conditions of $A$ to those of  $B$. Then $A/B$ is the Lorentzian continuation of the swap wormhole. All told, the formula for the entanglement entropy of universe $A$ turns out to be  
\be
S(\rho_{A}) = \underset{I}{\text{min ext}} \left[ \f{{\rm Area} (A/B, \p I)}{4G_{N}}  + S_\text{eff} ( I ) \right] \, , 
\ee 
where ${\rm Area} (A/B, \p I)$ denotes the area of $\p I$ in the new spacetime $A/B$ and $S_\text{eff}$ is the bulk entanglemement entropy  on $I$ of the effective matter theory in $A/B$.

Our swap wormhole is reminiscent of the bra-ket wormhole discussed in \cite{Chen:2020tes}.  A bra-ket wormhole connects a bra and ket which would na\"ively be created by disjoint gravitational path integrals.  A swap wormhole also connects gravitational boundary conditions which would not na\"ively be joined; namely a bra state in a copy of universe $A$ can join to a ket state in a copy of universe $B$.  
A swap wormhole can also be regarded as a Euclidean version of the Polchinski-Strominger  wormhole \cite{Polchinski:1994zs} which connects black hole interiors, see  \cite{Hsin:2020mfa,Marolf:2020rpm} for recent related studies. Swap wormholes help to maximize the action by making the bra and the ket state in every universe identical. The resulting entanglement entropy is given by the formula similar to \eqref{eq:islandformula} but the extremization must be  taken on the geometry constructed by suitably gluing   universes.

Our setup is also related to the process of black hole evaporation as follows. Let us regard the Hilbert space of one of the universes as a model of radiation degrees of freedom in the  evaporation process, and  let the degrees of freedom on the other universe correspond to the radiating black hole.  The emergence of the new spacetime $A/B$ in which the two gravitating  universes $A$ and $B$ are glued in our setup should  have a counterpart in the actual black hole evaporation process, suggesting the existence of a wormhole between the black hole and the early Hawking radiation in a manner reminiscent of \cite{Maldacena:2013xja}. In Euclidean signature this represents a virtual wormhole, but after continuation to Lorentzian signature, our results show that when the entanglement is large there is a phase where two disjoint microscopic Hilbert spaces may develop an effective description in semiclassical gravity that involves an Einstein-Rosen bridge between naively disconnected universes.

Six sections follow.  In Sec.~\ref{sec:setup} we explain the technical setup of two disjoint universes, and review our previous results for the entanglement entropy when one of the universes  is non-gravitating.
In Sec.~\ref{sec:cylinder-quotient} we describe the zoo of wormholes that appears, and why the cylinder wormhole between universes effectively reduces to the swap wormhole in some limits. 
In Sec.~\ref{sec:swapwormhole}, we show that the swap wormhole dominates the entanglement entropy in the large entanglement limit.    In Sec.~\ref{sec:twogravitating} we use these results to calculate the entanglement between two gravitating universes.   Finally, in Sec.~\ref{sec:lorentzian} we discuss the Lorentzian interpretation of these results, and the location of the ``island'' that contributes to the generalized entropy formula.  We conclude in Sec.~\ref{sec:disc} with a discussion of implications of our results and future directions.

While this paper was in preparation we received \cite{Anderson:2021vof} which has some overlap with our work.

\section{Setup}\label{sec:setup}

Consider two   disjoint gravitating universes, A and B. These  universes cannot  communicate classically but  may be  entangled quantum mechanically.   We then define  a conformal  field theory on  each universe, say $CFT_{A}$ and $CFT_{B}$, chosen to be identical for simplicity, and turn on semiclassical JT gravity  on both A and B. Furthermore, we assume the cosmological constant is negative, so that in Euclidean signature the geometry of both universes is the same hyperbolic disk. However, we allow the  universes to have different dilaton profiles. These are  technical assumptions, and   generalizations  to other cases are possible, but may require the inclusion of cosmic domain walls separating regions with different physical properties.

Thus, the total effective action on each universe is\footnote{ We will follow the notation of \cite{Balasubramanian:2020coy,Balasubramanian:2020xqf}.}
\be
\log Z= \log Z_{{\rm CFT}} - \f{\phi_{0}}{4\pi} \left[ \int_{D} R +\int_{\p D} 2K \right] - \int \f{\Phi}{4\pi} (R-\Lambda) - \f{\Phi_{b}}{4\pi} \int 2K. \label{eq:action}
\ee
 Since the  universes are disjoint, the total  Hilbert space is  a tensor product $\mathcal{H}_{A} \otimes \mathcal{H}_{B}$.  On this bipartite system, we define the thermofield double state
\be
 |\Psi \ra=\sum_{i} \s{p_{i}} |\psi_{i} \ra_{A} \otimes  |\psi_{i} \ra_{B}, \quad p_{i} =\f{e^{-\beta E_{i}}}{Z(\beta)} ,
 \label{eq:tfdstate}
\ee
where the states $\ket{\psi_i}$ are defined in the microscopic UV-complete theories so that they create bulk CFT states with energy $E_{i}$ by insertion of an operator $\psi_i$ on the south pole of a half-disk.
Furthermore, with this definition, the inner product expression $\inner{\psi_i}{\psi_j}$ represents a circular boundary condition for the gravitational path integral with the operator $\psi_i$ inserted at the north pole and $\psi_j$ inserted at the south pole.
If we fill in this circular boundary with a disk, the CFT path integral on that geometry will compute an overlap of CFT energy eigenstates.
For now, we leave the metric boundary conditions unspecified, but we have in mind the boundary conditions for a Euclidean black hole (whose inverse temperature might not be equal to $\beta$), as we will later see.
We will refer to $\beta$ as the entanglement temperature in \eqref{eq:tfdstate} to distinguish it from the temperature of a black hole.

\begin{figure}[t]
    \centering
    \includegraphics[scale=.25]{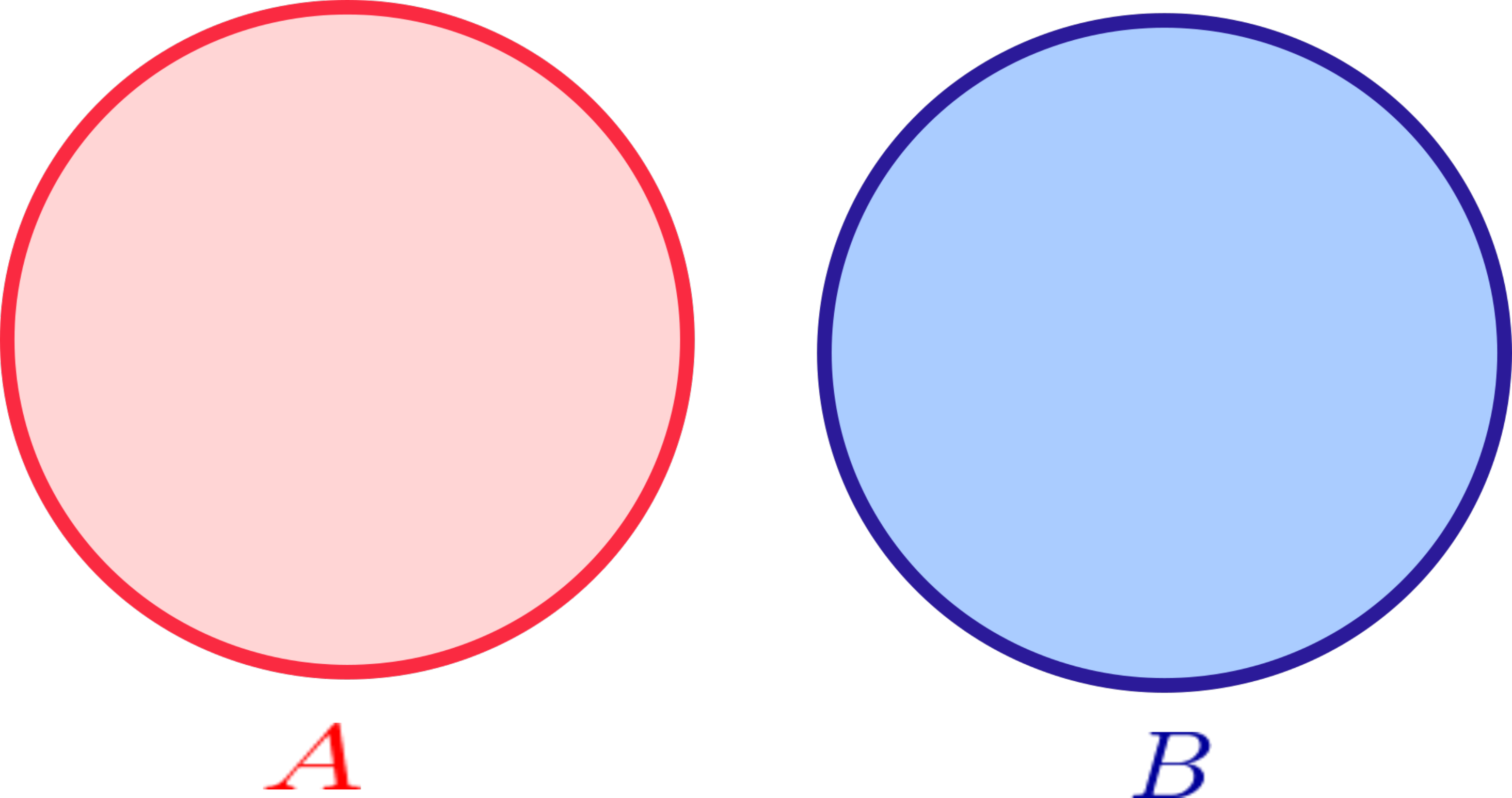}
    \caption{\small{We have two universes $A$ and $B$ with semiclassical Euclidean disk solutions.}}
    \label{fig:Setup}
\end{figure}

We are interested in the entanglement entropy between the two  gravitating universes. Since they are disjoint, it is possible to define such an entanglement entropy even in the presence  of gravity on each universe in terms of  the von Neuman entropy of the reduced density matrix $\rho_{A}$ on  universe $A$:
\be
S(\rho_{A}) =- \tr \rho_{A} \log \rho_{A}, \quad \rho_{A} =\tr_{B} |\Psi \ra \la \Psi |. 
\ee
We can compute this by using the replica trick,
\be 
S(\rho_{A}) = \lim_{n \rightarrow1 } \; \f{1}{1-n} \log \tr \rho_{A} ^{n}.
\label{eq:eedef}
\ee
From the definition we have  
\be
\tr \rho_{A} ^{n}  =\f{1}{Z_{1}^{n}}\sum_{i_{k}, j_{k}=1} ^{\infty} \prod^{n}_{k=1}\s{p_{i_{k}} p_{j_{k}}}\; 
 \la \psi_{i_{k}}|\psi_{j_{k}}\ra_{A_{k}}\;  \la \psi_{j_{k}}|\psi_{i_{k+1}}\ra_{B_{k}} \, \label{eq:renyithiscase}
\ee
where $Z_{1}$ is a normalization factor defined by
\be
Z_{1} = \sum_{i,j} \s{p_{i}p_{j}} \;\la \psi_{i} | \psi_{j} \ra_{A}\la \psi_{j} | \psi_{i} \ra_{B}
\ee
and we denote the $k^\text{th}$ copies of universes $A$ and $B$ by  $\{A_{k}, B_{k} \}$.   We will sometimes refer to this pair as the $k^\text{th}$ replica, because it is analogous to the replica sheet of a R\'enyi entropy calculation in quantum field theory. But we should keep in mind that in our setting, each replica consists of two gravitational boundary conditions associated with $A_k$ and $B_k$.

\vspace{0.2cm}

Each individual overlap $\la \psi_{i} | \psi_{j} \ra$ receives its dominant contribution (up to non-perturbative higher genus gravitational corrections which will not concern us in this work) from a path integral on the disk. This is because the CFT sector of the excited state $|\psi_{j} \ra$ on the reflection symmetric slice of the disk can be prepared by a Euclidean path integral on the lower half of the disk with the insertion of a local operator $\psi_{j}(0)$ at the south pole.  This local operator is related to $ |\psi_{j} \ra$ by the state-operator correspondence of conformal field theory. Similarly, the CFT sector of $\la \psi_{i} |$ can be prepared by a path integral on the upper upper half of the disk, with an insertion $\psi_{i}(\infty )$ at the north pole. Combining these, the overlap is related to the two point function of such local operators inserted at the north and the south pole of the disk, $\la \psi_{i} | \psi_{j} \ra = \la \psi_{i} (\infty)\psi_{j} (0) \ra $.

\vspace{0.2cm}

More generally, the right hand side of \eqref{eq:renyithiscase} can be computed by a gravitational path integral involving $2n$ circular boundary conditions ($n$ associated with universe $A$ and $n$ with universe $B$).  We will evaluate the path integral in the semiclassical limit $G_{N} \rightarrow 0$, by picking up appropriate saddle points of various topologies.  In general there is no established  prescription to properly choose such saddles. As shown in  \cite{Penington:2019kki,Almheiri:2019qdq}, in order to avoid information loss, it is necessary to include wormholes connecting replica copies of the universes in the gravitational path integral.

\vspace{0.2cm}

\subsection*{One gravitating universe}

If we turn off gravity on universe $A$, the states are orthogonal, i.e., $\la \psi_{j_{k}}|\psi_{i_{k+1}}\ra_{A_{k}}= \delta_{j_{k} i_{k+1}}$,  so  \eqref{eq:renyithiscase} reduces to
\be
\tr \rho_{A} ^{n}  =\f{1}{Z_{1}^{n}}\sum_{i_{k}, j_{k}=1} ^{\infty} \prod^{n}_{k=1}p_{i_{k}} \; 
 \la \psi_{i_{k}}|\psi_{i_{k+1}}\ra_{B_{k}}. 
 \ee
We computed this gravitational path integral in \cite{Balasubramanian:2020coy,Balasubramanian:2020xqf} by including a replica wormhole connecting $n$ copies of the gravitating universe $\{B_{k}\}_{k=1}^{n}$.  The resulting entanglement entropy was given by the island formula, 
\be 
 S(\rho_{A})= {\rm min} \{S_{{\rm no-island}}, S_{{\rm island}} \}\, ,
\ee
where $S_{{\rm no-island}}$ is the thermal entropy of the CFT with inverse temperature $\beta$, which is the analogue in this setting of Hawking's result for the entropy of black hole radiation. $S_{{\rm island}}$ is coming from the contribution of the replica wormhole, and is given by the generalized entropy, 
\be
S_{{\rm island}}= \underset{\overline{C}}{\text{min ext}} \left[\Phi [\p\overline{C}] +S_{\beta} [\overline{C}]  -S_{{\rm vac}} [\overline{C}] \right], \label{eq:iscbar}
\ee
where $\overline{C}$ is an interval on the gravitating universe $B$, and $S_{\beta} [\overline{C}] $ is the CFT entanglement entropy of the subregion $\overline{C}$ in the thermofield double (TFD) state. Similarly, $ S_{{\rm vac}} [\overline{C}] $ is the vacuum entropy on  the same subregion. Since  this TFD state is pure on AB, it is instructive to write it as
\be 
S_{{\rm island}}= \underset{C}{\text{min ext}} \left[ \Phi [\p AC] +S_{\beta} [AC]  -S_{{\rm vac}} [AC] \right].
\ee
In this expression, the region $C$ in the gravitating universe is identified as the entanglement island. In practical calculations, the expression \eqref{eq:iscbar} is more convenient.


We will briefly explain  how $S_{\beta} [\overline{C}]  -S_{{\rm vac}} [\overline{C}] $ part appears in  $S_\text{island}$, as this will be important for our later discussions.
Practically, the contribution of the replica wormhole  to the gravitational path integral  can be evaluated as follows. First, we  first introduce a cut  $C$ on each copy of the disk universe $B_{k}$, and glue all $n$ copies along the cut $C$. Denote the resulting branched disk by $\Sigma_n(C)$. Then, on the replica wormhole, the required product of  overlaps is  computed by a correlation function on the branched disk:
\be
\prod^{n}_{k=1} \; 
 \la \psi_{i_{k}}|\psi_{i_{k+1}}\ra_{B_{k}}  = \la \prod ^{n}_{k=1}   \psi_{i_{k}} (\infty_{k}) \psi_{i_{k+1}} (0_{k+1})
\ra_{\Sigma_{n}(C)},
\ee
where $\infty_{k}$ and $0_k$ denote the north and south poles of the $k^\text{th}$ disk, respectively.
By performing the index sums and using a CFT identity involving the cut \cite{Balasubramanian:2020coy}, we find
\be
\sum_{i_{k}} ^{\infty} \prod^{n}_{k=1}p_{i_{k}}  \la \psi_{i_{k}}|\psi_{i_{k+1}}\ra_{B_{k}}   = e^{-I_{{\rm grav}}[\Sigma_{n}(C)]}\;  \f{\tr  \rho_{\beta, \overline{C}}^{n}}{\tr  \rho_{{\rm vac},\overline{C}}^{n}} \, ,
\ee
where $\tr \rho_{\beta, \overline{C}}^{n}$ is the CFT R\'enyi entropy of the canonical ensemble $\rho_{\beta} =\frac{1}{Z} e^{-\beta H}$ on the subregion $\overline{C}$ and $\tr \rho_{{\rm vac},\overline{C}}^{n}$ is the 
 R\'enyi entropy of the vacuum state on the same subregion.  The quantity $I_{{\rm grav}}[\Sigma_{n}(C)]$ denotes the gravitational action of the branched disk $\Sigma_{n}(C)$ with the understanding that the contributions of the conical singularities are subtracted. 
By taking the $ n \rightarrow 1$ limit, we get the expression for $S_{{\rm island}}$ in \eqref{eq:iscbar}.

\vspace{0.2cm}

\section{Replica wormholes between two universes}\label{sec:cylinder-quotient}

The basic idea behind replica wormholes, as explained in  \cite{Penington:2019kki,Almheiri:2019qdq}, is that the  gravitational path integral which computes R\'{e}nyi entropies is allowed to include wormholes between replica copies of a gravitating theory.   In our case, we have two gravitating theories A and B.  Thus, when we evaluate the Euclidean path integral to compute  \eqref{eq:renyithiscase}, we are led to consider saddlepoints with no wormholes, wormholes just between copies of A,  wormholes just between copies of B, and saddlepoints with copies of A and of B separately connected by wormholes.   We are going to propose that if A and B are both gravitating, we should treat them democratically and additionally permit wormholes between copies of A and copies of B.  Indeed, the results in later sections show that such wormholes between the different universes are necessary to achieve a unitary Page curve in the present setting.
Our quantitative analysis uses the two-dimensional theory \eqref{eq:action}, and we therefore use  two-dimensional terminology throughout to discuss topology as well, but there are no real dimensional restrictions at the qualitative level.
In other words, our discussion could be lifted to $D$ dimensions by fibering a transverse sphere $S^{D-2}$ over all the topologies in question, though the actual calculations in that setting would be prohibitively difficult owing to the relative difficulty of higher dimensional gravity compared to the JT theory.

\subsection{New contributions modifying the island formula}

We wish to compute the R\'{e}nyi entropy \eqref{eq:renyithiscase} using semiclassical gravity coupled to a quantum conformal field theory on both universes $A$ and $B$.
In order to do so, we must introduce a gravitational boundary condition for each inner product appearing in that expression, and subsequently compute the gravitational actions and quantum matter field partition functions of all geometries and topologies which obey those boundary conditions.
For the $n^\text{th}$ R\'{e}nyi entropy, we will have $2n$ boundary conditions: $n$ associated with universe $A$ and $n$ associated with universe $B$.

We must sum over a number of gravitational configurations.
For instance, we might decide to fill in each boundary condition separately with a disk, and compute the path integral on $2n$ disconnected disks.
Famously, after analytic continuation in $n$ this choice leads to an entropy $S(\rho_A)$ approximately equal to the entropy of quantum matter fields that are entangled between universes $A$ and $B$.
Such an entropy is unbounded, as the temperature of the matter fields which enters in the definition of the state \eqref{eq:tfdstate} can be arbitrarily large.
This is essentially Hawking's answer for the entropy of Hawking radiation.
However, as we have learned, there are topologically nontrivial gravitational configurations involving Euclidean wormholes that connect some subsets of the gravitational boundary conditions, and these configurations (if we choose to include them in the path integral) can come to dominate the R\'{e}nyi entropy \eqref{eq:renyithiscase}.
If we choose the disconnected disk configuration for all universe $A$ boundary conditions, the $n$-boundary genus zero wormhole connecting all boundary conditions of universe $B$ eventually dominates the R\'{e}nyi entropy calculation, and this configuration leads to the \ island formula for the microscopic entanglement entropy of a non-gravitating subregion (universe $A$, in this context) \cite{Balasubramanian:2020coy}. 

In our current situation, we have dynamical gravity on both universes $A$ and $B$.
So, in the calculation of the R\'{e}nyi entropies \eqref{eq:renyithiscase}, we will encounter at least the saddles which lead to both the quantum matter field entropy and the island formula relative to universe $B$.
However, it is possible that the island formula will produce an incorrect answer for the entropy when both universes are gravitating.
This is because, in the presence of dynamical gravity on both universes $A$ and $B$, there is an additional class of Euclidean wormholes we may allow ourselves to include in the path integral: wormholes which connect boundary conditions associated with universe $A$ to boundary conditions associated with universe $B$.

Though it is not obvious that we should allow such connections between different universes, our  perspective is democratic: all gravitational boundary conditions are created equal.
So, regardless of what universe a given boundary condition is associated with, we shall allow Euclidean wormhole configurations which connect it to any other boundary condition.
Following the same logic which leads to the island formula, we are led to consider a fully-connected $2n$-boundary genus zero wormhole $M_{2n}$ which connects all $2n$ boundary conditions appearing in \eqref{eq:renyithiscase}.
After a short classification of other saddles, the contribution of this fully-connected wormhole to the R\'enyi entropy will be the main focus of our CFT and JT gravity analysis in the following subsection.

\subsection{A zoo of wormholes}

\begin{figure}
\centering
\includegraphics[width=6cm]{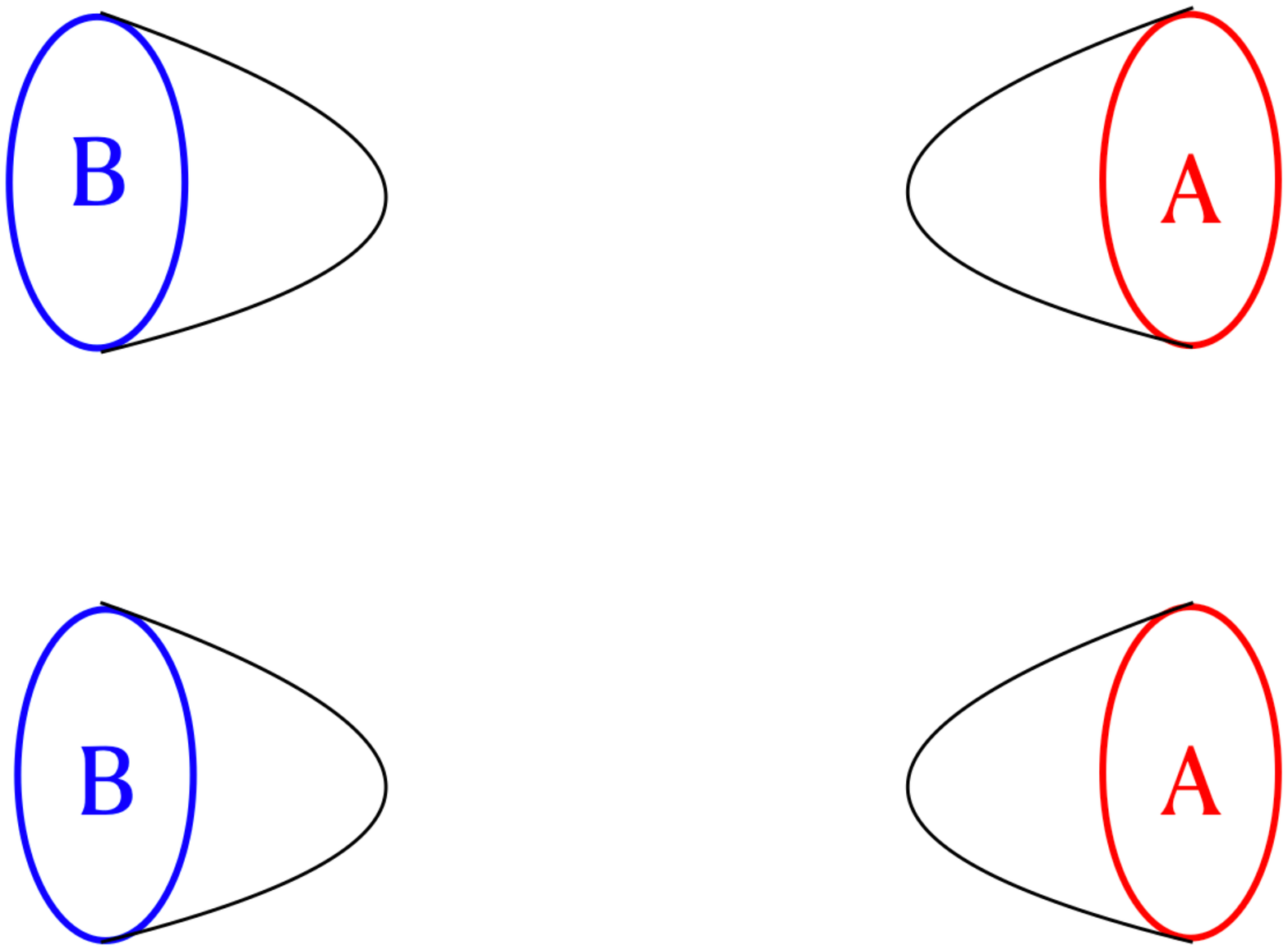}
\hspace{1 cm}
\includegraphics[width=6cm]{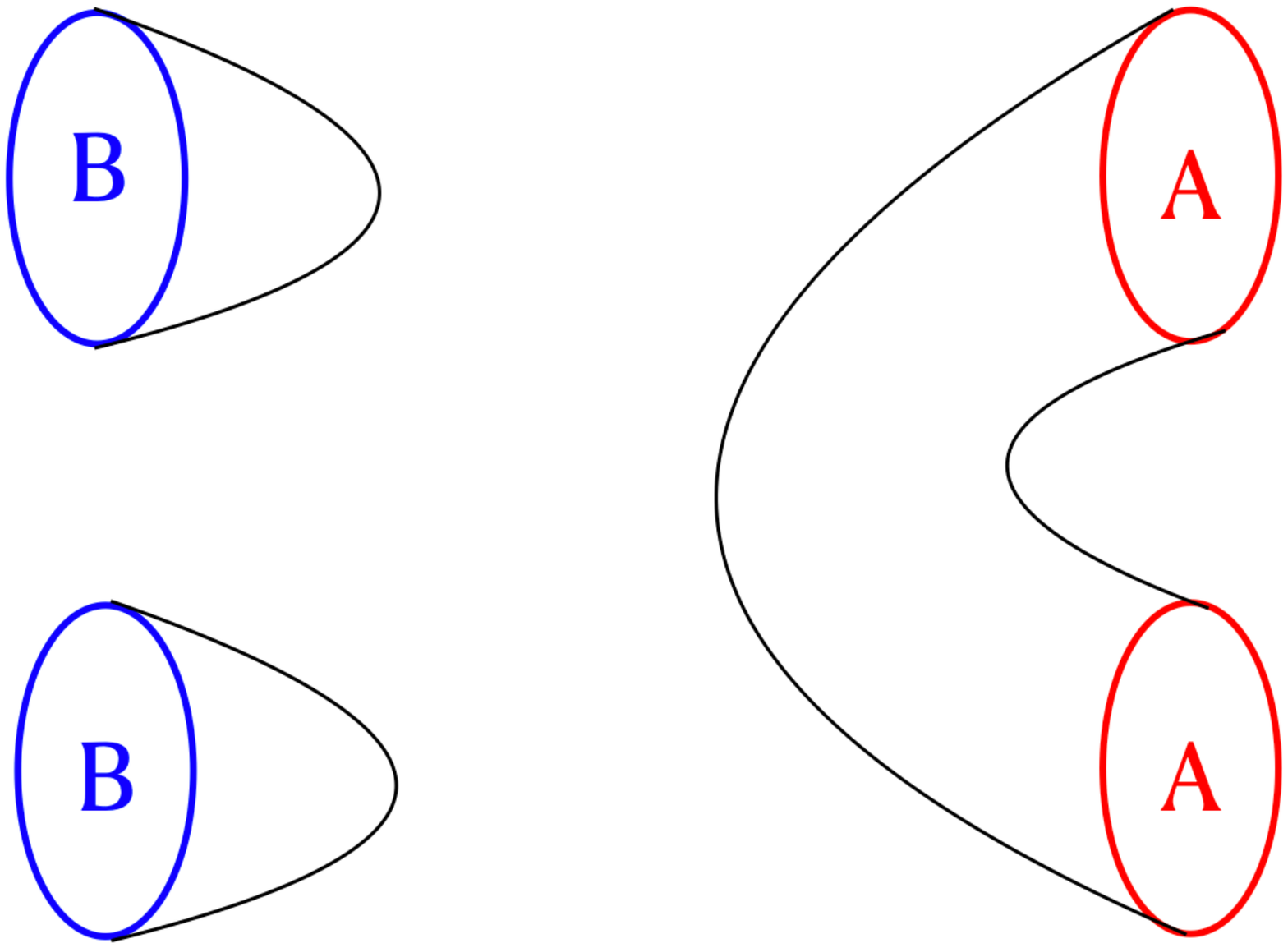}
\hspace{1 cm}
\includegraphics[width=6cm]{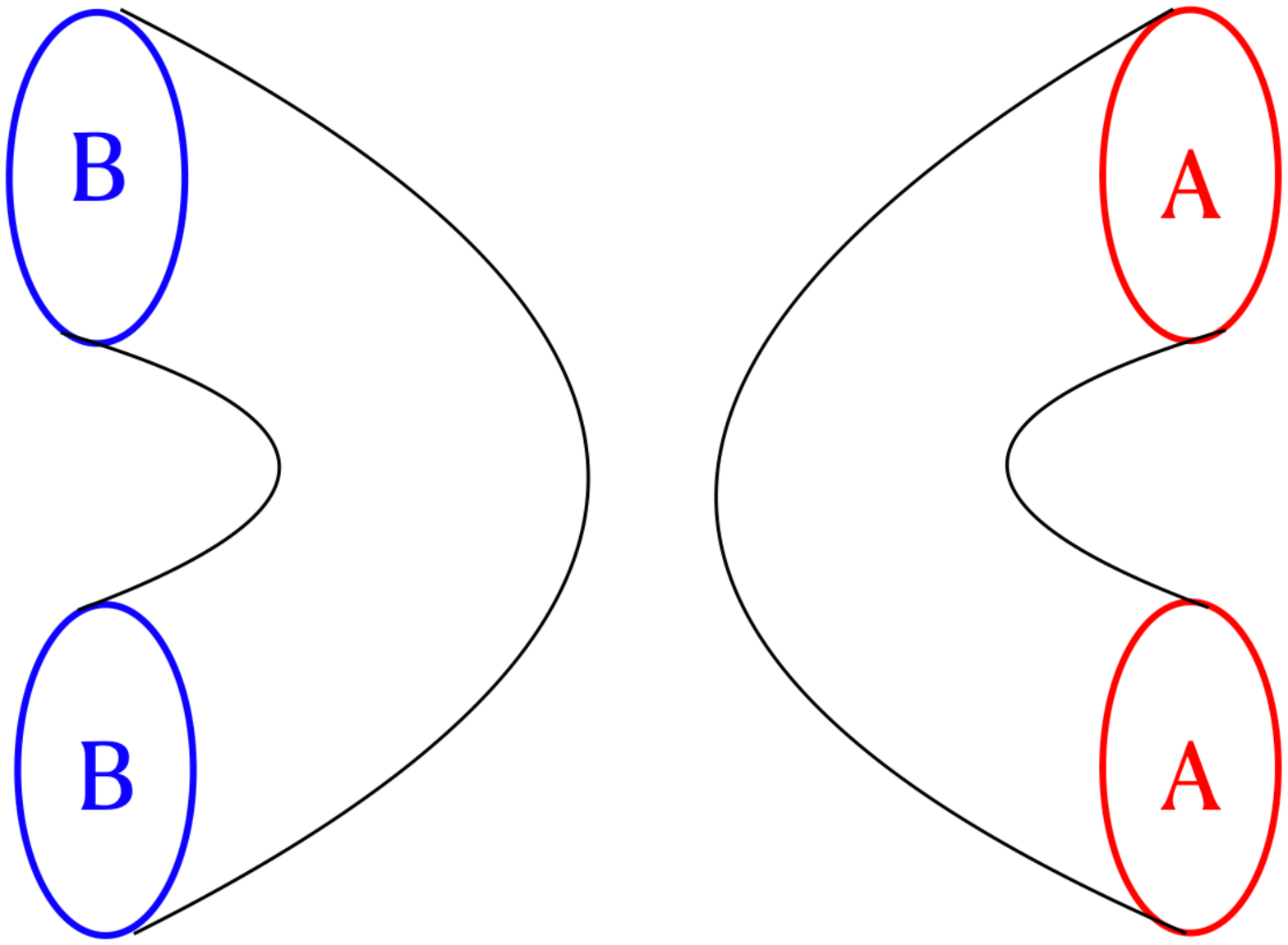}
\hspace{1 cm}
\includegraphics[width=6cm]{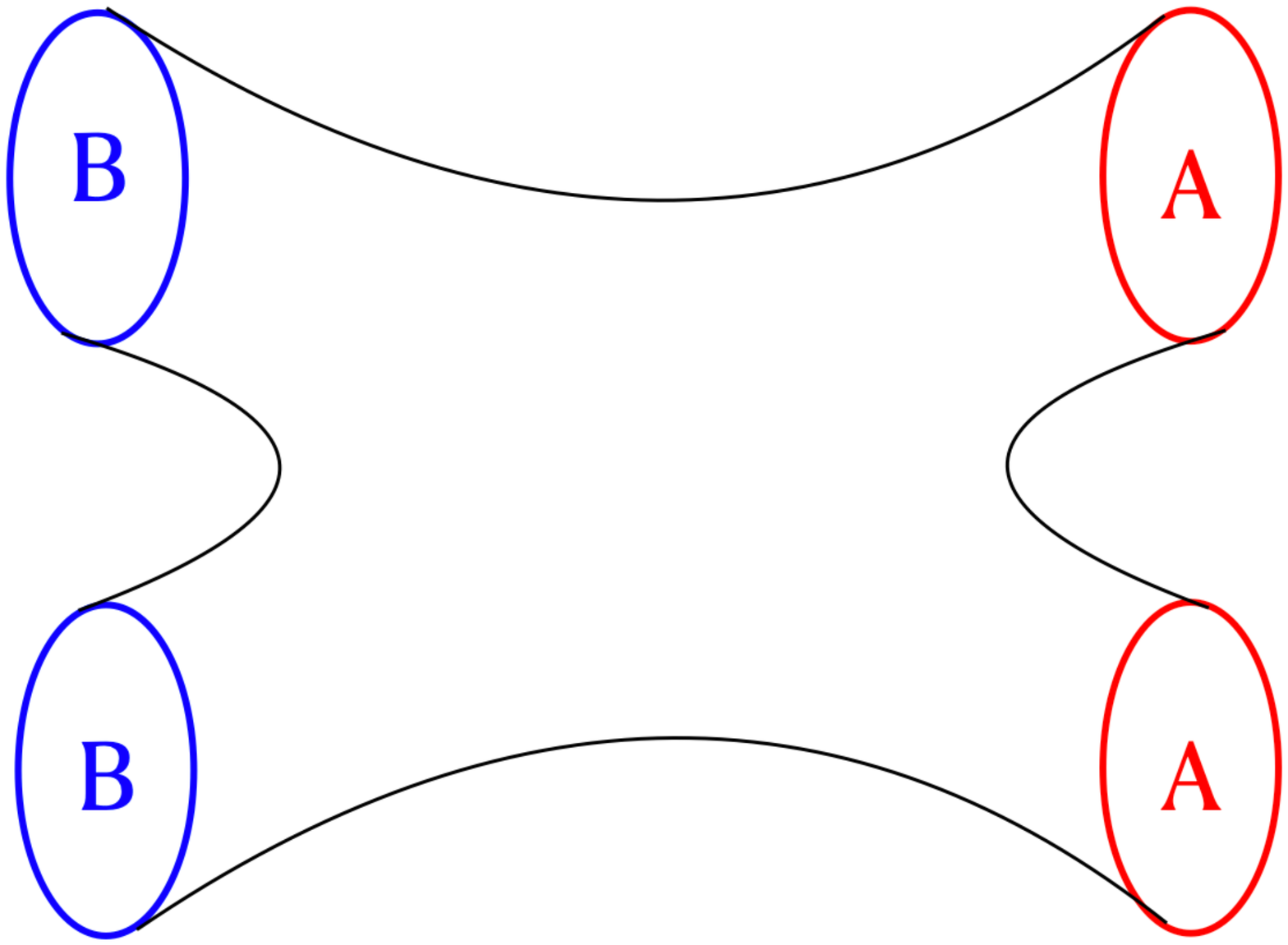}
\caption{\small{Possible gravitational configurations connecting copies of  the universes $A$ and $B$ in the $n=2$ R\'{e}nyi entropy \eqref{eq:renyithiscase}.  (Top left) Type I configuration where all copies are disconnected.     (Top right) Type II$_{A}$ configuration where all copies of the universe $A$ are connected by a replica wormhole.  (Bottom left) Type III configuration where  all copies of $A$ are connected by a replica wormhole, and all copies of $B$ are connected by another replica wormhole, but these two wormholes are not connected.   (Bottom right) Type IV configuration where all copies are connected by a single wormhole.}}
\label{fig:wormholes}
\end{figure}

In JT gravity theory coupled to a quantum matter CFT as in \eqref{eq:action}, we can employ the replica trick to calculate the various contributions to \eqref{eq:renyithiscase} in more detail.
In \cite{Balasubramanian:2020coy}, this procedure was carried out for situations where only one of $A$ or $B$ is gravitating, and led to a version of the island formula.
Here there are a variety of contributions to consider, even if we assume some form of replica symmetry exists for the bulk topology, so we will proceed systematically through the possible configurations (see Fig. \ref{fig:wormholes}):
\begin{enumerate}[align=left]
     \item[Type I:] The fully disconnected  configuration of $2n$ disks.
     \item[Type II$_A$:]
     All replica copies of universe $A$ are connected by an $n$-boundary genus zero replica wormhole $R_{A}$, but all  copies of universe $B$ are still totally disconnected.
     \item[Type II$_{B}$:]
     All replica copies of universe $B$ are connected by an $n$-boundary genus zero replica wormhole  $R_{B}$, but all  copies of universe $A$ are still totally disconnected.
     \item[Type III:] All replica copies of universe $A$ are connected by an $n$-boundary genus zero replica wormhole $R_{A}$, and all replica copies of universe $B$ are connected by another $n$-boundary genus zero replica wormhole  $R_{B}$, but these two replica wormholes are not connected. 
     \item[Type IV:] All copies of universes $A$ and $B$ are connected by a single $2n$-boundary genus zero wormhole. 
 \end{enumerate}
To see which configuration gives the dominant contribution,  let us evaluate the contribution of each saddle to the gravitational path integral.

\subsubsection*{Type I}

The semiclassical partition function of the type I configuration is straightforward to evaluate.  
In this configuration, the overlaps are  orthogonal $\la \psi_{i} | \psi_{j} \ra _{A}= \la \psi_{i} | \psi_{j} \ra _{B}=\delta_{ij}$, which leads to
\be
Z_{\text{type I}} = Z_{{\rm grav}} [A]^{n}Z_{{\rm grav}} [B]^{n} Z_{{\rm CFT}} (n \beta).
\ee
The quantity $Z_\text{grav}[A]$ is the partition function of quantum gravity on the disk topology for universe $A$, and is evaluated semiclassically as $e^{-I_\text{grav}[A]}$.

\subsubsection*{Type II}
Similarly, we can evaluate the semiclassical partition function of the type II configurations.
This is almost identical to the evaluation in \cite{Balasubramanian:2020coy}, where one of the universes is non-gravitating.
The off-shell action on the replica wormhole can be evaluated at non-integer $n$ by the standard Lewkowycz-Maldacena trick \cite{Lewkowycz:2013nqa}. 
We quotient by the $\mathbb{Z}_n$ replica symmetry by defining $\Sigma_{n} = M_{n}/\mathbb{Z}_{n}$ and subtracting the contribution of the orbifold singularity.
This quotient can be identified with  a branched disk, constructed by  gluing $n$ copies of the disk along a cut $C$. The  location of the cut  is determined by  solving gravitational equations of motion.
The CFT partition function $Z_\text{CFT}$ is also affected by the wormhole connection, namely the product of the overlaps is replaced by a $2n$-point correlation function 
\be
\prod_{k=1}^{n} \la \psi_{i_{k}} |\psi_{j_{k}}\ra_{A_{k}} =\la \prod^{n}_{k=1}\psi_{i_{k}} (\infty_{i_{k}})\psi_{j_{k}} (0_{j_{k}}) \ra_{\Sigma_{n}} .
\ee
Then the full semiclassical type II$_A$ partition function is (and an analogous result holds for type II$_B$) 
\be
Z_{\text{type II}_A} = \left( Z_{{\rm CFT}} (\beta)\; Z_{{\rm grav}} [B]\right)^{n} \left(\f{\tr \rho_{\beta,\overline{C}}^{n}}{\tr \rho_{{\rm vac},\overline{C}}^{n}} \; Z_{{\rm grav}} [\Sigma_{n}]\right) .
\ee
The result of the type II$_A$ and II$_B$ wormholes look almost identical to those obtained in the derivation of the island formula, but we note one distinction which also affects type I wormholes.
Since gravity is technically turned on in both universes, the normalization factor $Z_1^n$ which enters the definition of the R\'enyi entropy will itself receive contributions from wormholes.
In certain quantities, this effect can be large \cite{Engelhardt:2020qpv}, and we will see in Sec. \ref{sec:swapwormhole} that this effect is also important in our setup, and indeed is partially responsible for the sub-dominance of type II$_A$ and II$_B$ wormholes at large temperatures $\beta \to 0$. 

\subsubsection*{Type III}
We now turn to the type III configuration. 
Unlike the previous configurations, the CFT partition function on this topology does not have a nice R\'{e}nyi entropy interpretation because the operator configuration does not produce a product of pure state density matrices. Instead, we can only write it as a product of correlation functions
\be
\prod_{k=1}^{n} \la \psi_{i_{k}} |\psi_{j_{k}}\ra_{A_{k}} \la \psi_{j_{k}} |\psi_{i_{k+1}}\ra_{B_{k}} = \la \prod^{n}_{k=1}\psi_{i_{k}} (\infty_{A_{k}})\psi_{j_{k}} (0_{A_{k}}) \ra_{\Sigma_{n}}\la \prod^{n}_{k=1}\psi_{j_{k}} (\infty_{B_{k}})\psi_{i_{k+1}} (0_{B_{k+1}}) \ra_{\Sigma_{n}} .
\ee
Here $\infty_{A_{k}}$ and  $0_{A_{k}}$ denotes the north and the south poles of the $k^\text{th}$ replica copy of the disk universe $A_{k}$, and we used a similar notation $\infty_{B_{k}}$ and  $0_{B_{k}}$  for the universe $B_{k}$.
This leads to a rather complicated, though still in principle calculable, expression for $Z_\text{type III}$:
\begin{equation} 
\begin{split}
Z_\text{type III} & = Z_{{\rm grav}} [\Sigma_{n}[A]]Z_{{\rm grav}} [\Sigma_{n}[B]] \\
& \quad \times \sum_{i_{k},j_{k}} \s{p_{i_{k}} p_{j_{k}}} \la \prod^{n}_{k=1}\psi_{i_{k}} (\infty_{A_{k}})\psi_{j_{k}} (0_{A_{k}}) \ra_{\Sigma_{n}}\la \prod^{n}_{k=1}\psi_{j_{k}} (\infty_{B_{k}})\psi_{i_{k+1}} (0_{B_{k+1}}) \ra_{\Sigma_{n}} .
\end{split}
\label{eq:ztypeIII}
\end{equation}

\subsubsection*{Type IV}
\begin{figure}
    \centering
    \includegraphics[width=7cm]{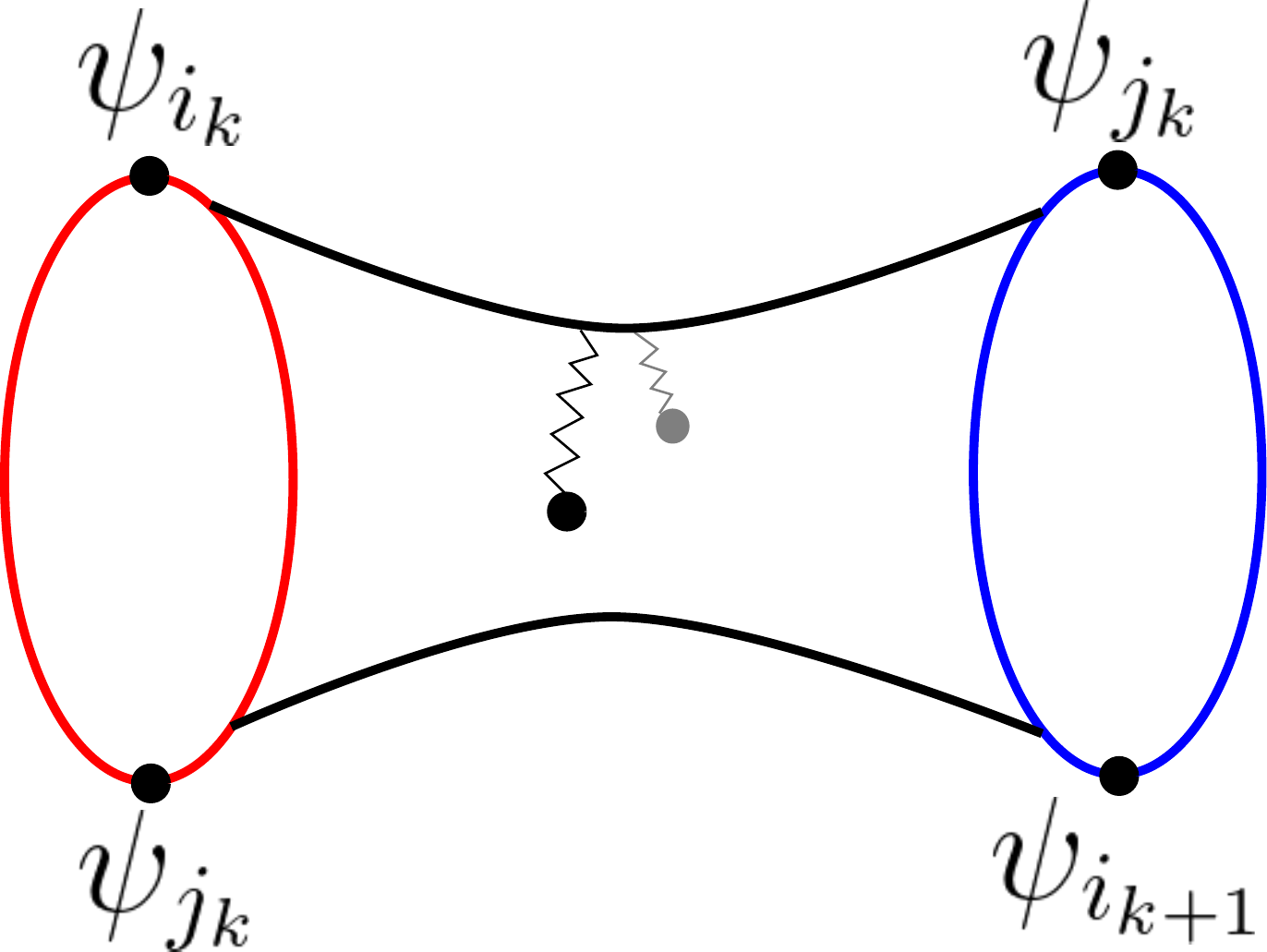}
    \caption{\small{The $n$-sheeted cylinder manifold $\Sigma^{\text{cyl}}_n$ with operator insertions appearing in \eqref{eq:cylindercorrl}.  Going through the cut takes us to a different sheet, on which the particular operator insertions are different. The cut here is in grey compared to Fig.~\ref{fig:replica} because here we have only deformed the manifold $M_{2n}$ slightly, as opposed to taking a quotient of the topology by a $\mathbb{Z}_n$ replica symmetry.  Indeed, with the operator insertions, this path integral on its own is not actually replica symmetric, and one of our main aims in Sec.~\ref{sec:effective} is to start with the CFT path integral on this branched manifold and manipulate it into a nicer form more amenable to analytic continuation in $n$. }}
    \label{fig:cylinder-operators}
\end{figure}

Finally, we turn to the fully-connected wormhole $M_{2n}$ that joins all of the gravitational boundary conditions appearing in \eqref{eq:renyithiscase}. 
An immediate novel feature to which we will later return is that a quotient of the geometry $M_{2n}$ by $\mathbb{Z}_n$ leads to a cylinder topology.
The analysis of this wormhole will lead to a formula for entropy which involves a Cauchy surface $\Sigma(AB)$ (containing the quantum extremal surface $\partial C$) that is \textit{not} of the standard product form $\Sigma(A) \times \Sigma(B)$, as occurs in the usual island formula.
We begin by splitting the calculation into the gravity part and the CFT part, 
\be
Z[M_{2n}] = Z_{{\rm grav}}[M_{2n}]\;Z_{{\rm CFT}}[M_{2n}] .
\label{eq:partitionM2n}
\ee
We would like to evaluate the CFT partition function for the fully connected wormhole.
This is equivalent to evaluating the following correlation function on the branched cylinder $\Sigma^{{\rm cyl}}_{n}$
\be
Z_{{\rm CFT}}[M_{2n}]= \la \prod^{n}_{k=1} \psi_{i_{k}} (\infty_{A_{k}}) \psi_{j_{k}} (0_{A_{k}})  \psi_{j_{k}} (\infty_{B_{k}}) \psi_{i_{k+1}} (0_{B_{k}})\ra_{\Sigma^{{\rm cyl}}_{n}} .
\label{eq:cylindercorrl}
\ee
Here, $\infty_{A_{k}}$ is understood as the top left corner of the cylinder, and $0_{A_{k}}$ is the bottom left.
That is to say, $\infty_{A_k}$ is the north pole of the left circular boundary and $0_{A_k}$ is the south pole of that same boundary (Fig.~\ref{fig:cylinder-operators}).
Similarly, $\infty_{B_{k}}$ and $0_{B_{k}}$ are the top and the bottom of the right circular boundary.  We will argue below that this wormhole dominates the computation of the entropy.

When only one of the universes is non-gravitating, the above product of overlaps on the replica manifold $R_A$ or $R_B$ does have a nice R\'enyi entropy interpretation because it reduces to a product of pure state density matrices, and this fact was crucial for obtaining the island formula in \cite{Balasubramanian:2020coy}.
Therefore, in this context, one may wonder if this correlation function has a nice interpretation in terms of a R\'enyi entropy on the cylinder.
In general, it is hard to find such an interpretation because the above pattern of operator insertions on the cylinder cannot easily be written as a product of pure state density matrices, as in the case of the disk \cite{Balasubramanian:2020coy}.
An even further complication for the cylinder is that the state-operator correspondence is not applicable.
Below, we will attempt to give a R\'enyi entropy interpretation to the correlation function in the high entanglement temperature $\beta \to 0$ limit, but we emphasize before proceeding that the correlation function \eqref{eq:cylindercorrl} gives an exact accounting of the CFT contribution to the entropy on this topology.
Thus, it represents an intermediate regime where the gravitational sector is geometric but the bulk entropy contribution is not of the usual R\'enyi form.
It would be interesting to study the correlation function \eqref{eq:cylindercorrl} in more detail from a CFT perspective, but we leave this for future work.

\subsection{Dominance of the cylinder wormhole}
\label{sec:dominance}
In the high temperature limit $\beta \rightarrow 0$ the path integral is dominated by the type IV configuration where all copies of the universes are connected by a wormhole. Intuitively, this must occur because at very high entanglement temperatures $\beta$ in the state \eqref{eq:tfdstate}, there are two copies of any given field theory operator $\psi_{i_k}$ (similarly for $\psi_{j_k}$), and this pair of operators must be connected by a short path running through the Euclidean bulk in order for the on-shell action to be minimized. 

In more detail, in the CFT sector, any given operator always appears twice. 
For example, $\psi_{j_{k}}$  is inserted at the the south pole of $A_{k}$ by the $\psi_{j_{k}}(0_{A_{k}})$ operator and at the north pole of $B_{k}$ by the $\psi_{j_{k}}(\infty_{B_{k}})$ operator. 
Of course, this is just because the state $ | \psi_{j_{k}} \ra $ appears twice in the R\'enyi entropy \eqref{eq:renyithiscase}. 
Then, in order to maximize the CFT partition function, the saddle must satisfy the following condition. 
For given operator $\psi_{j_{k}}$, the operator nearest to it is the other insertion of $\psi_{j_{k}}$ itself.
As we will see shortly, if this condition is satisfied, one can take the OPE limit  $\psi_{j_{k}}(0_{A_{k}}) \rightarrow \psi_{j_{k}}(\infty_{B_{k}})$  in the high temperature limit $\beta \rightarrow 0$, and subsequently one can indeed check that this configuration dominates.
Coming back to \eqref{eq:ztypeIII}, we see that the above condition is not satisfied because one insertion of $\psi_{j_{k}}$ is living in universe $A_{k}$ and the other insertion is in universe $B_{k}$, but there is no sense in which these insertions can become close because the two universes are not even connected. 
Thus, in the type I, II and III wormholes,  $\psi_{i_{k}}(\infty_{B_{k}})$ cannot be the nearest operator to $\psi_{i_{k}}(0_{A_{k}})$. Instead, this condition is only satisfied when all gravitational boundary conditions for both universes $A$ and $B$ are connected by a single wormhole, as in the type IV wormholes.

This is simply a field-theoretic version of the index-sum phenomenon in the popular end-of-the-world brane model, which leads to similar entanglement-driven transitions \cite{Penington:2019npb,Balasubramanian:2020hfs,Balasubramanian:2020jhl,Anderson:2021vof}.
It also appears explicitly in computations which probe finer aspects of the gravitational ensemble \cite{Stanford:2020wkf}.

\subsection{Effective description of the cylinder wormhole}\label{sec:effective}

To find an effective R\'enyi entropy interpretation of \eqref{eq:cylindercorrl}, we take the limit $\beta \to 0$ in \eqref{eq:tfdstate}.
In this limit, matching CFT operator insertions must be connected in pairs by short (in renormalized length) paths through the Euclidean bulk in order to minimize the on-shell action as required by the equations of motion.
In the fully connected type IV wormhole, this is certainly possible, and we may think of the manifold $M_{2n}$ with a particular set of operator insertions as an $n$-fold cover of a cylinder with two operators on each boundary, where these operators change on different sheets.
One of these sheets is depicted on the left in Fig.~\ref{fig:degenerationlimit}.
The moduli of this branched cylinder (including the locations of the branch points) will be fixed by the extremization condition.
Subsequently, we will find an effective description of the path integral in the CFT sector by taking a suitable OPE limit and rewriting a long Euclidean evolution as a vacuum projector.

\subsubsection*{Moduli extremization}
\begin{figure}
\centering
\includegraphics[width=7cm]{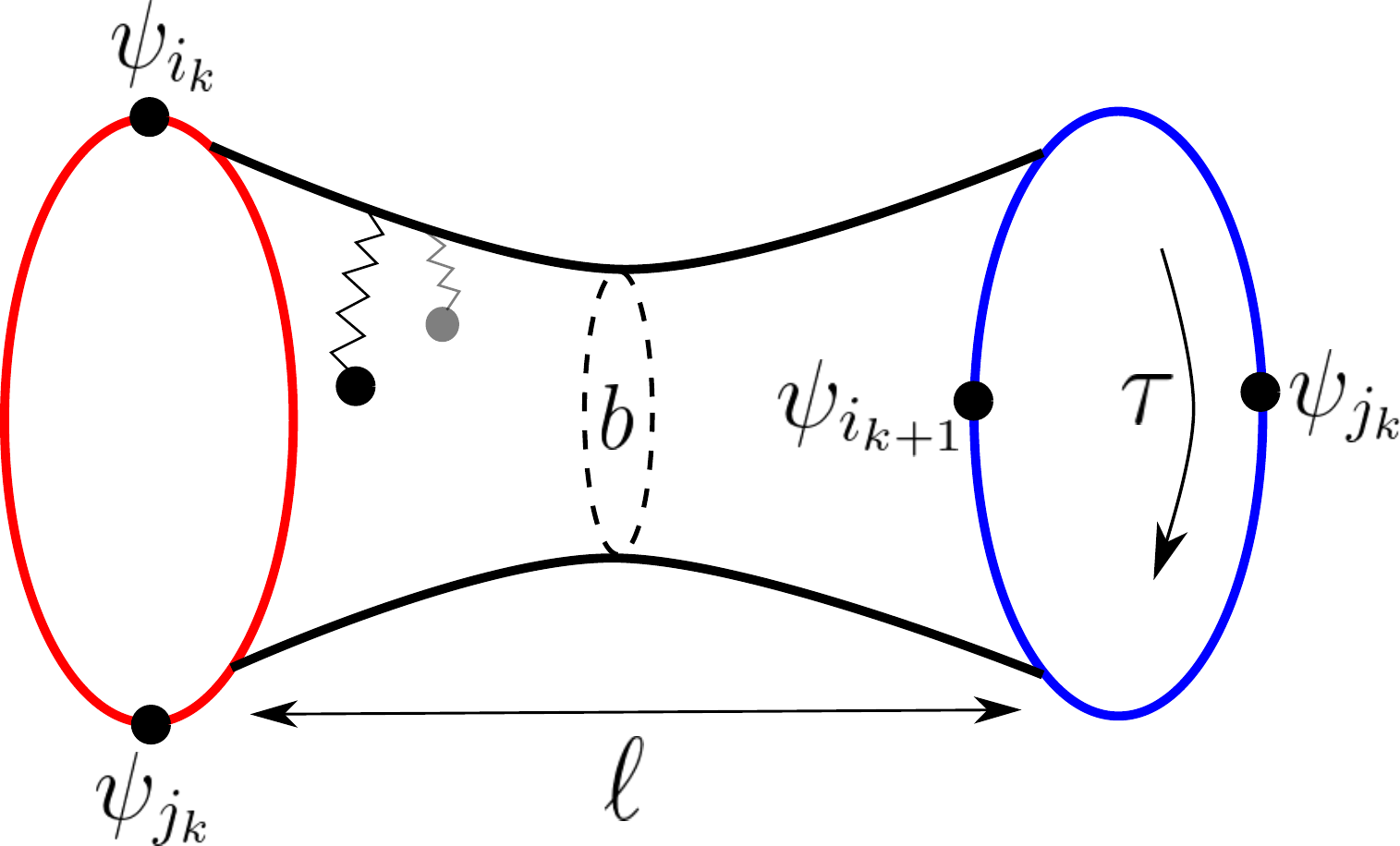}
\caption{\small{Two moduli of the cylinder wormhole $b$ and $\tau$. The twist parameter $\tau$ rotates universe $B$, and the circumference $b$ measures the width of the wormhole.  The twist changes the distance between $\psi_{j_{k}} (\infty_{A})$ in universe $A$ and $\psi_{j_{k}} (0_{B})$ in universe $B$ in  the correlator \eqref{eq:cylindercorrl}. This distance is minimized when $\tau = \pi$.  We can also think of the renormalized length $\ell$ between two boundary points, which becomes small when the circumference $b$ is large.} }
\label{fig:moduli}
\end{figure}

The cylinder itself has two moduli: the twist $\tau$ that describes the relative rotation between the two boundaries, and the minimal geodesic circumference $b$.  In pure JT gravity, the only effect of $\tau$ is to contribute a factor of $b$ to the measure on moduli space \cite{Saad:2019lba} because the cylinder is invariant under a twist of one side.  However, in our case the operator insertions in \eqref{eq:cylindercorrl} lead to a different on-shell action and CFT partition function for different values of $\tau$, so we must extremize over $\tau$. Minimizing the renormalized length $\ell$ between matching $\psi_{j_k}$  insertions (one on the boundary of the universe $A$, i.e., the red circle in Fig. \ref{fig:moduli}, and the other on the boundary of universe $B$, i.e., the blue circle) in \eqref{eq:cylindercorrl} then leads a selection of $\tau = \pi$ so that one half of the cylinder is rotated and the operators $\psi_{j_k}$ are aligned across the wormhole.

To reduce the renormalized length $\ell$ between the operator insertions, the asymptotic boundaries of cylinder must be brought closer together. Since the metric on the cylinder has a constant negative curvature, it can be thought of as two trumpet geometries \cite{Saad:2019lba} (one asymptotic boundary and one circular geodesic boundary) glued together along their geodesic boundaries which have length $b$.  Thus when the asymptotic boundaries come closer, the gluing surface will increase in circumference.\footnote{In JT gravity, one often uses a basis $\ket{\ell}$ of definite renormalized length states for the Hilbert space on an interval \cite{Harlow:2018tqv}.  Here we are treating gravity semiclassically, so the metric on the geometry is fixed by equations of motion, and we will have definite values for the moduli $b$ and $\tau$ as well as the renormalized length between any two boundary points.}  So bringing the asymptotic boundaries of the cylinder closer ($\ell \rightarrow 0$) is equivalent taking the modulus $b$ to be large (Fig.~\ref{fig:moduli}).  As a side comment, in pure JT gravity there is no saddle on the cylinder topology \cite{Saad:2019lba,Maldacena:2018lmt}, so it is really the large backreaction of the CFT fields which supports a large $b$ wormhole saddle in our situation.

\begin{figure}
\centering
\includegraphics[width=12cm]{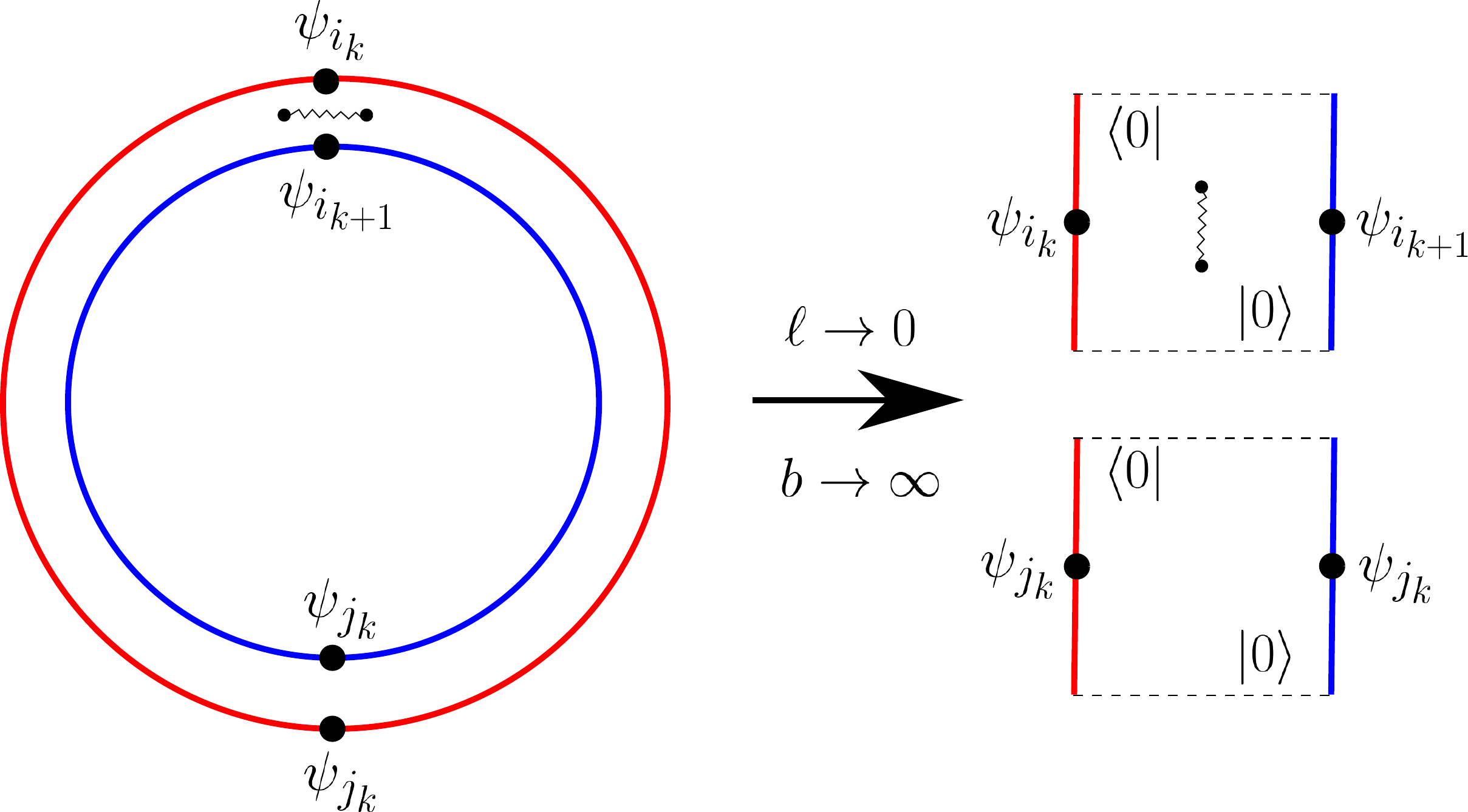}
\caption{\small{Left: The short length or large circumference limit of the wormhole connecting the universes $A$ and $B$ where the cylinder has been drawn as an annulus and we have already set $\tau = \pi$ to rotate the $A$ boundary relative to the $B$ boundary. 
The black dots represent operator insertions on the cylinder wormhole, and we have drawn the cut in its extremized location between $\psi_{i_k}$ and $\psi_{i_{k+1}}$. 
Right:  When  $\ell \rightarrow 0$ or $b\to\infty$, the long Euclidean evolutions on the annulus may be replaced with CFT vacuum projectors.  We have exaggerated the Euclidean evolution on the right.}}
\label{fig:degenerationlimit}
\end{figure}

Having brought the operators $\psi_{j_k}$  in \eqref{eq:cylindercorrl} close together by a choice of $\tau$ and $b$, we now turn to the pair of $\psi_{i_k}$.
The short path between this pair must pass once through the branch cut, since a single sheet (as depicted in Fig.~\ref{fig:degenerationlimit}) does not contain a pair of $\psi_{i_k}$. For instance, in the correlation function \eqref{eq:cylindercorrl} on the branched cylinder,  $\psi_{i_{k}}(\infty_{A_{k}})$ is in the $k^\text{th}$ sheet, whereas $\psi_{i_{k}}(0_{B_{k-1}})$ is in the $(k-1)^\text{th}$ sheet. 
To minimize the renormalized length between this pair, the location of the cut enters the game.
We see that by moving the cut directly in between the $\psi_{i_k}$ and $\psi_{i_{k+1}}$ insertions on Fig.~\ref{fig:degenerationlimit} so that it is parallel to the asymptotic boundaries, we can minimize the lengths between pairs of $\psi_{i_k}$ and also pairs of $\psi_{i_{k+1}}$.

\subsubsection*{OPE limit, vacuum projection, and state-operator correspondence}

As we have discussed, the $\beta \to 0$ limit of \eqref{eq:tfdstate} results in a R\'enyi entropy calculation that involves a CFT path integral on a large circumference cylinder, or equivalently a short annulus as depicted in Fig.~\ref{fig:degenerationlimit}.
There is a cut between the operators $\psi_{i_k}$ and $\psi_{i_{k+1}}$ which connects the $n$ sheets, each of which have distinct operator insertions.
The short annulus also results in the two $\psi_{j_k}$ insertions approaching each other as the renormalized length becomes small.
Since these identical operators are close together and well-separated from the $\psi_{i_k}$ operators, we can consider an OPE limit where they fuse to become the identity operator. 
Therefore, the correlation \eqref{eq:cylindercorrl} is significantly simplified in this limit. 
We will see that this simplification is crucial for recovering a formula for the entanglement entropy which involves an expression like the generalized entropy when both of the universes are gravitating. 

 \vspace{0.2cm}

Specifically, we now argue that the annulus on which we compute the path integral is effectively split into two disks in the limit $\ell \rightarrow 0$. 
This provides a practical description of the cylinder wormhole and the correlator \eqref{eq:cylindercorrl} which we will make use of in the later sections. 
To see this, let us study the $\ell \rightarrow 0 $ limit from a slightly different perspective. 
We start from the path integral form of \eqref{eq:cylindercorrl} on the annulus.
Since the annulus has a periodic direction, we can write the path integral as 
a trace with insertions of local operators.
Taking a time coordinate $t$ along the periodic direction, the path integral is
\be
\tr \left[ U(\beta_{{\rm an}}, \beta_{{\rm an}}/2)  \; \psi_{i_{k}}  \psi_{i_{k+1}} \; U( \beta_{{\rm an}}/2, 0 )\; \psi_{j_{k}}  \psi_{j_{k}} \right].
\label{eq: traceform}
\ee
$U(t_{a}, t_{b})$  is a Euclidean time evolution operator  on the  $S^{1}$ direction and $\beta_{{\rm an}}$ is the period of this $S^1$.
Of course, the full path integral for \eqref{eq:cylindercorrl} involves all sheets joined through the cut. 
In \eqref{eq: traceform} we omit these features to simplify the presentation.
 
 \vspace{0.2cm}
 
When the size of the interval Cauchy slice $I$ is small, $|I|=\ell\rightarrow 0$, then in the rescaled coordinates with fixed width $|I| = \pi$ the circumference of the annulus becomes large $\beta_{{\rm an}} \sim 1/\ell \rightarrow \infty$. 
In this limit, the Euclidean time evolution operator becomes a projection to the ground state $| 0 \ra $ on the interval 
\be
 U( \beta_{{\rm an}}/2, 0 )  \sim  e^{-\beta_{{\rm an}} E_{0}/2} |0 \ra \la 0|,
\ee
where $E_{0}$ denotes the ground state energy.
As a consequence, the path integral on the annulus \eqref{eq: traceform} is factorized into two 2-point functions
\be
\tr \left[ U(\beta_{{\rm an}}, \beta_{{\rm an}}/2)  \; \psi_{i_{k}}  \psi_{i_{k+1}} \; U( \beta_{{\rm an}}/2, 0 )\; \psi_{j_{k}}  \psi_{j_{k}} \right]  \rightarrow  \la 0 |\; \psi_{i_{k}}  \psi_{i_{k+1}}|0 \ra \la 0 |\psi_{j_{k}}  \psi_{j_{k}} |0 \ra, \quad  \ell \rightarrow 0,\label{eq:corrfactrized}
\ee
as drawn in Fig.~\ref{fig:degenerationlimit}.
The CFT vacuum on an interval $\ket{0}$ is created by a half-disk path integral.
Our situation here is quite similar, though there is a slightly unorthodox aspect which is that the half-disk in question has ``mixed" boundary conditions.
That is to say, one side of the half-disk has boundary conditions associated with universe $A$, and the other side has those of universe $B$.
If these were the same, the state-operator correspondence would allow us to generate the state $\ket{0}$ by simply inserting the identity operator at the boundary.
Here we insert an effective boundary condition-changing operator, or ``swaperator", which is meant to transition the boundary conditions from those of universe $A$ to those of universe $B$.
This operator does not affect the CFT sector, and its only effect on the gravitational sector is to change the asymptotic boundary conditions appropriately.

\begin{figure}
    \centering
    \includegraphics[width=12cm]{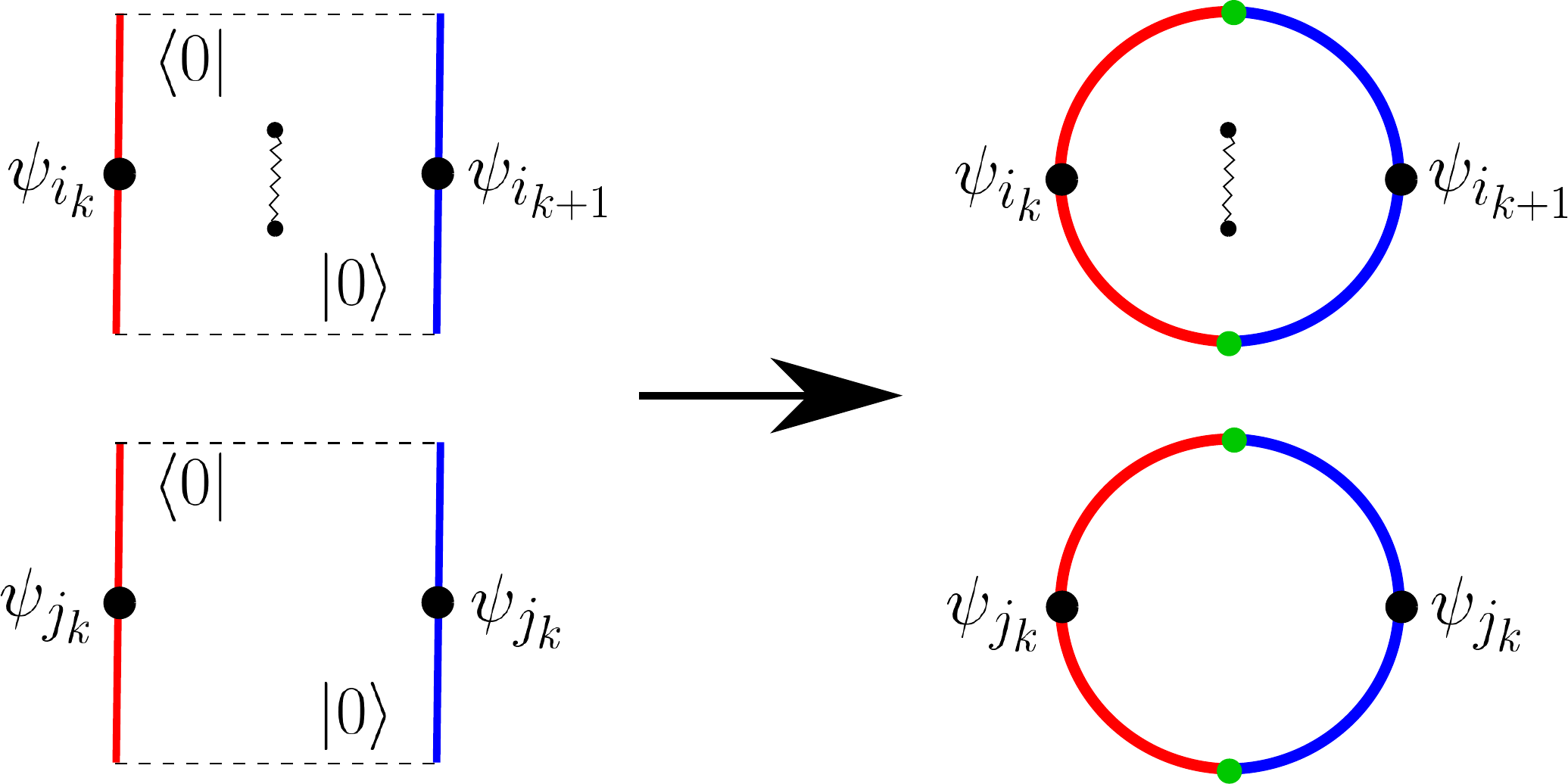}
    \caption{The state-operator correspondence allows us to replace the CFT vacuum state with a path integral on a half-disk.  The green dots are swaperators which change the gravitational boundary conditions from those of universe $A$ to those of universe $B$, but these do not affect the CFT sector.}
    \label{fig:state-operator}
\end{figure}

With this caveat about swaperators, the two correlation functions on the right hand side of \eqref{eq:corrfactrized}  can be written as CFT path integrals on disks with the two operators inserted at the endpoints of the $\mathbb{Z}_2$-symmetric slice (Fig.~\ref{fig:state-operator}).
Then, our original correlation function \eqref{eq:cylindercorrl} becomes 
\begin{equation} 
Z_{{\rm CFT}}[M_{2n}] \rightarrow  \; \la  \prod^{n}_{k=1} \psi_{i_{k}} (\infty_{A_{k}}) \psi_{i_{k+1}} (0_{B_{k}})\ra_{\Sigma^{{\rm Disk}}_{n}} \times  \prod_{k=1}^n \avg{\psi_{j_k}(\infty_{B_k}) \psi_{j_k}(0_{A_k}) }_{\text{Disk}} .
\label{eq:cylindercorr}
\end{equation}
Here by $\la \cdots \ra_{{\rm Disk}}$
we mean the correlation function evaluated on the disk, and $\la \cdots \ra_{\Sigma^{{\rm Disk}}_{n}}$ is the correlation function evaluated on the branched disk, and in both cases we must include swaperators at the junctions where the $A$ boundary meets the $B$ boundary.
Again, we stress that the disk now is not the original disk of universe $A$ or $B$, but a new one constructed by gluing the upper half of universe $A$ and the lower half of universe $B$, as shown in Fig. \ref{fig:state-operator}.

\subsubsection*{Gravitational sector}
Thus far, we have focused on the CFT sector of the calculation, and we found that the correlation function \eqref{eq:cylindercorrl} has an effective description in terms of a disk and branched disk geometry where half of the disk's boundary is associated with universe $A$'s boundary conditions, and the other half is associated with those of universe $B$.

\begin{figure}
    \centering
    \includegraphics[scale=.45]{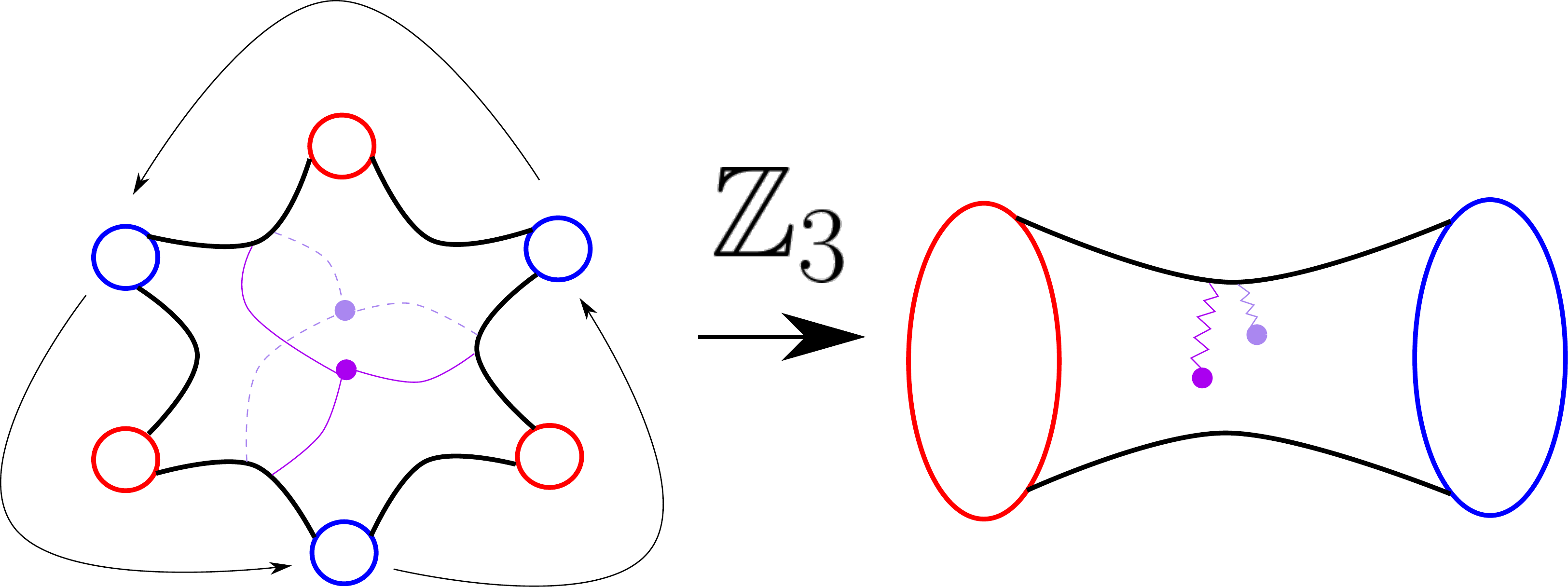}
    \caption{\small{The $M_6$ geometry (without operator insertions) has a $\mathbb{Z}_3$ cyclic replica symmetry which sends replicas of universe $A$ (red) to other replicas of universe $A$, and similarly for $B$ (blue).  After quotienting by the action of this symmetry \cite{Lewkowycz:2013nqa}, a cylinder is produced with one $A$ and one $B$ boundary.  The two fixed points of the $\mathbb{Z}_3$ action on $M_6$ (purple and light purple dots) map to the two cosmic brane and twist operator points on the quotient cylinder.  The curves connecting the fixed points on $M_6$ which are exchanged under the $\mathbb{Z}_3$ action are mapped to a cut along which we obtain a field theory contribution.}}
    \label{fig:replica}
\end{figure}

We now turn to the gravity term: the partition function $Z_{{\rm grav}}[M_{2n}]$ in \eqref{eq:partitionM2n}.
By following the standard prescription \cite{Lewkowycz:2013nqa}, analytic continuation in $n$ of this object can be achieved by computing the on-shell action of its quotient $\Sigma_{n} \equiv M_{2n}/\mathbb{Z}_{n}$, which is the cylinder wormhole of Fig.~\ref{fig:replica}, together with a cut $C$ joining two cosmic branes (and twist operators).
In the $n \rightarrow 1$ limit, we obtain 
\be 
Z_{{\rm grav}}[M_{2n}] =(Z_{{\rm grav},1})^{n} e^{-(n-1) \Phi^{{\rm cyl}}_{n=1}[\p C]},
\label{eq:zgrav}
\ee
where $Z_{{\rm grav},1}$ is the gravitational partition function (meant to be evaluated semiclassically) of the $n=1$ cylinder wormhole geometry, and $\Phi^{{\rm cyl}}_{n=1}[\p C]$ is the dilaton on the cylinder evaluated at the end points of the cut $\p C$.  
Notice that in the actual value of the R\'enyi entropy, $Z_{{\rm grav},1}$ does not show up, because the R\'enyi entropy is a ratio $Z[M_{2n}]/Z_{1}^{n}$.
A similar phenomenon occurs in other derivations of gravitational entropy formulas, where cosmic branes contribute an area term that depends only on the local background metric, and the full gravitational action of the solution does not appear.

 \vspace{0.2cm}

We saw that in the $\ell \rightarrow 0$ limit, the CFT partition function $Z_{{\rm CFT }} [M_{2n}]$ had an effective description in terms of a disk constructed by gluing two universes.
We may also wonder whether or not a similar effective description is available for the gravitational partition function $Z_{{\rm  grav}} [M_{2n}]$.  
At first sight, this seems problematic, because the topologies of the cylinder and the two disks are different.
The Euler characteristic of each manifold  appears in the topological term of the JT gravity action, so the gravitational action of the cylinder differs from that of the disk even at $O(1/G_{N})$. However, as we have discussed, the topological piece does not show up in the actual R\'enyi entropy in the $n\rightarrow 1$ limit and instead this limit is solely determined by the dilaton profile on the $n=1$ geometry.
So, if we can show the dilaton profile on the cylinder $\Phi^{{\rm cyl}}_{n=1}$ in \eqref{eq:zgrav} approaches the dilaton profile on the new disk $\Phi^{{\rm disk}}_{n=1} $ in the  $\ell \rightarrow 0$ limit, then we may conclude that there is an alternative description for the gravitational sector as well. 

\vspace{0.2cm}

The dilaton profile $\Phi^{{\rm cyl}}_{n=1}$ on the cylinder (or equivalently on the annulus) is fixed by solving the equation of motion of the dilaton, with the boundary conditions for universe $B$ on the inner circle of the annulus and those for universe $A$ on the outer circle. 
Again as in \eqref{eq:corrfactrized} and Fig.~\ref{fig:degenerationlimit}, when the width of the annulus is small, i.e., $\ell \rightarrow 0$, the annulus looks like two factorized long strips with appropriate boundary conditions, and thus $\Phi^{{\rm cyl}}_{n=1}$ is approximately equal (at least at the branch points) to the dilaton profile on the strip
\begin{equation} 
\Phi^{\text{cyl}}_{n=1} = \Phi^{{\rm strip}}_{n=1} .
\end{equation}
At a technical level, we now use a conformal mapping to argue for equivalence with the disk instead of replacing the long Euclidean evolutions by vacuum projectors as we did in the CFT sector.
Let $(t, \phi)$ be the coordinates on the (effectively infinite) strip, $ -\infty<t <\infty$ and $0<\phi <\pi$.   
We impose the boundary conditions for universe $A$ at $\phi =0$  and the conditions of  universe $B$ at $\phi =\pi$ on the strip. The metric on the strip reads
\be
ds^{2} = dt^{2} + d\phi^{2}.
\ee
This strip is conformally  equivalent  to a disk with  the coordinates $(\theta, \phi)$  and the metric
\be
ds^{2} = d\theta^2 + \sin^{2} \theta d\phi^{2}.
\ee
by the map $t=-1/\tan \theta$. This new disk can be regarded as the one constructed by suitably gluing two half-disks of universes $A$ and $B$ as in Fig.~\ref{fig:state-operator}. 
Specifically, on the new disk we impose the boundary conditions of universe $A$ on the left half of the boundary $\phi =0$ and those of universe $B$ on the right half.
Because the boundary conditions on the disk are precisely the boundary conditions of the strip passed through the conformal map, the solution to the dilaton equation on the disk will simply be the solution on the strip passed through the map.
Therefore, we have $\Phi^{{\rm cyl}}_{n=1} =\Phi^{{\rm disk}}_{n=1}$, where we mean to compare these two profiles after they have been mapped to the same manifold by the conformal map. 

 \vspace{0.2cm}

\section{Swap wormholes}
\label{sec:swapwormhole}

In the latter half of the previous section, we discussed a particular saddle for the R\'enyi entropy \eqref{eq:renyithiscase} where all universes are connected by a single wormhole.  
Taking its quotient by the replica symmetry $\mathbb{Z}_{n}$ led to the cylinder wormhole, which connects the two gravitating universe $A$ and $B$. 
We further argued that, when the length of the wormhole is small (which is expected to happen in the large entanglement limit between $A$ and $B$), there is an alternative description of the cylinder wormhole.   
Namely, the path integral on  the cylinder wormhole is factorized into two path integrals on two disks. 
This factorization occurred in both the gravitational sector where we explicitly analytically continued by a $\mathbb{Z}_n$ quotient and also in the CFT sector where we analyzed the full partition function using OPE limits, vacuum projection, and the state-operator correspondence.
On each disk, we impose the boundary conditions of universe $A$ on the upper half and those of universe $B$ on the lower half. 
Below, we will call these disks ``swap wormholes".  
In this section, we study the properties of swap wormholes and their contributions to the  gravitational path integral for the R\'enyi entropy \eqref{eq:renyithiscase}. 
In particular, we explain why this configuration becomes the dominant saddle in the high temperature (large entanglement) limit.

As a reminder, the end goal of introducing this effective swap wormhole picture for the cylinder wormhole contribution is to write the CFT contribution to the entanglement entropy of the state \eqref{eq:tfdstate} in the form of a bulk effective field theory entropy.
As we will now see, this leads to a formula for the overall entanglement entropy which is reminiscent of a generalized entropy, just like what occurs in the island formula.
However, as we remarked in the beginning of Sec.~\ref{sec:cylinder-quotient}, the generalized entropy which we will find is not computed on a Cauchy slice associated with the original slices $\Sigma(A)$ or $\Sigma(B)$, but rather exists in a combined slice $\Sigma(AB)$ that connects the two universes as in the ER = EPR proposal.

\vspace{0.2cm}

\begin{figure}
     \centering
     \includegraphics[scale=.2]{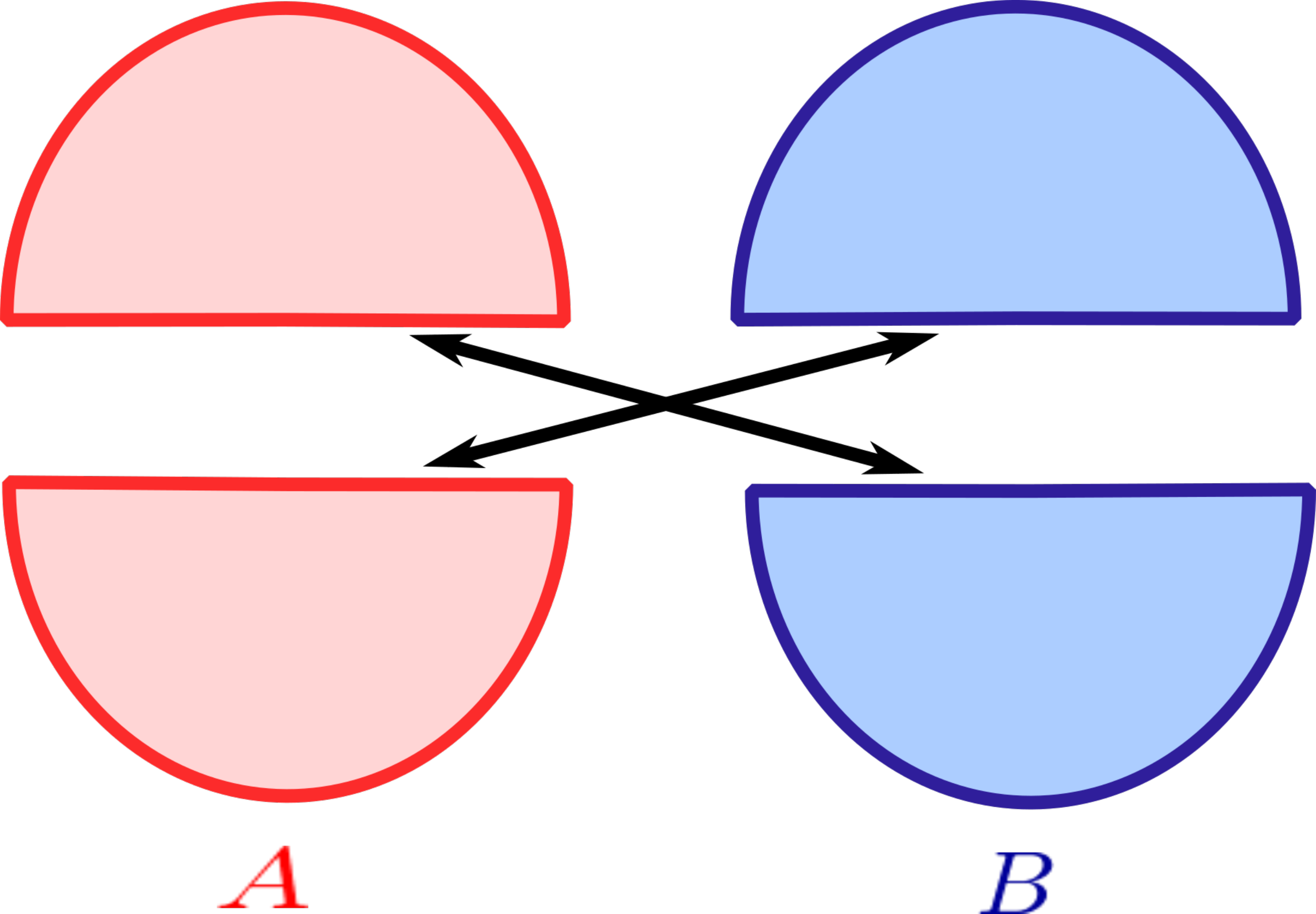}
     \hspace{3cm}\includegraphics[scale=.2]{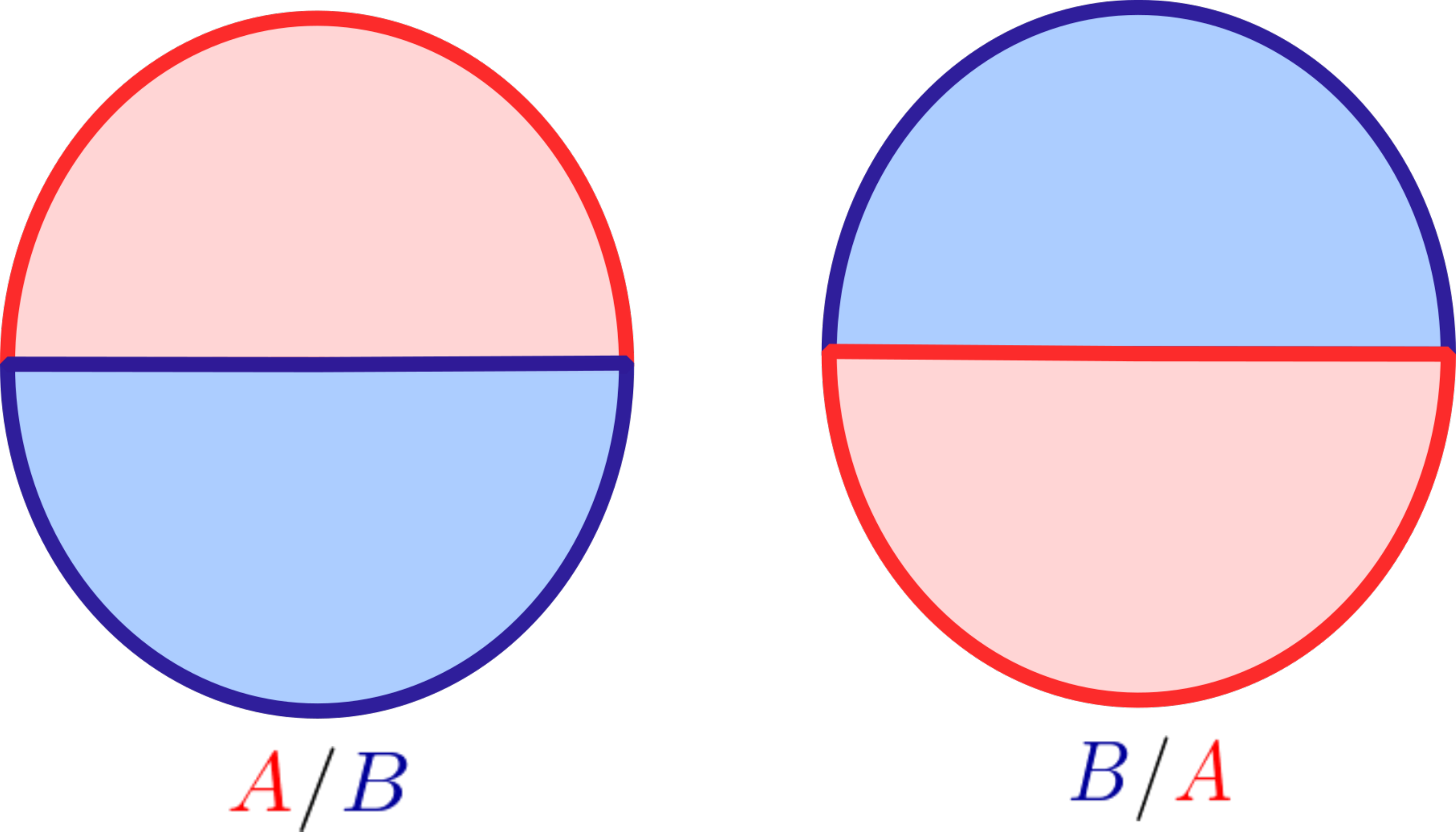}
     \caption{\small{Construction of a swap wormhole. 
     }}
     \label{fig:sphere-replicas}
 \end{figure}

In the swap wormhole picture of the replica manifold $M_{2n}$, we see from e.g. \eqref{eq:cylindercorr} that a swap wormhole in the $k^\text{th}$ replica connects the upper half of the $k^\text{th}$ disk universe $B$ to the lower half of the $k^\text{th}$ disk universe $A$ (Fig.~\ref{fig:sphere-replicas}). 
We denote the resulting manifolds constructed  in this way by $\{(A/B)_{k}, (B/A)_{k} \}$. 
In other words, $(A/B)_{k}$ is the manifold made by gluing the upper half of the disk $A_{k}$ with the lower half of the identical disk $B_{k}$ on the reflection symmetric slice. 
We similarly define  $(B/A)_{k}$ to be the another swap wormhole created by connecting the upper half of $B_{k}$ to the lower half of $A_{k}$.

  \vspace{0.2cm}

Since the metric is always fixed in JT gravity, we may split the disks $A_{k}$ and  $B_{k}$ and recombine them into the disks $(A/B)_{k}$ and $(B/A)_{k}$ without introducing any irregularities.
However, the dilaton profiles on the new configurations, which we denote $\Phi_{A/B}$ and $\Phi_{B/A}$, are different from the original dilaton profiles $\Phi_{A}$ and $\Phi_{B}$. 
$\Phi_{A/B}$ and $\Phi_{B/A}$ are constructed by solving the JT gravity equations of motion with the boundary conditions of universe $A$ on the upper half of the hyperbolic disk and those of universe $B$ on the lower half of the hyperbolic disk.
 
 \vspace{0.2cm}
 
Once we reconfigure the $k^\text{th}$ cylinder sheet on the replica manifold $M_{2n}$ into two swap wormholes, we have a description which essentially transforms the completely disconnected disks $\{A_{k}, B_{k} \}$ into
$\{(A/B)_{k}, (B/A)_{k} \}$ with a cut connecting the $(A/B)_k$, but with the $(B/A)_k$ all disconnected. 
The branched disk formed by the $(A/B)_k$ is just the one appearing in \eqref{eq:cylindercorr}.
In the disconnected saddle, we had $\psi_{i_{k}}$ at the north pole of $A_{k}$, and $\psi_{j_{k}}$ at the south pole.  
Similarly, we had $\psi_{j_{k}}$ and $\psi_{i_{k+1}}$ at the north and south pole of $B_{k}$.
Keeping track of the locations of these operators, we see that in the swap wormhole   $(A/B)_{k}$, the operator $\psi_{i_{k}}$ is inserted at the north pole and $\psi_{i_{k+1}}$ is inserted at the south pole of the disk.
Similarly, on $(B/A)_k$, we have $\psi_{j_{k}}$ at the north pole and $\psi_{j_k}$ at the south pole. 
This pattern of operator insertions matches that of the CFT correlator  \eqref{eq:cylindercorrl} on the cylinder wormhole in the the small length limit $\ell\rightarrow 0$, which appears in the gravitational path integral for the R\'enyi entropy \eqref{eq:renyithiscase}. 
Of course, this reflects the fact that the swap wormhole provides an effective description of the $\ell \rightarrow 0$ limit of the cylinder wormhole.

\subsection{Dominance of swap wormhole in high temperature limit} 
 
In order to illustrate how the swap wormhole dominates over the completely disconnected saddle in the gravitational path integral, we can consider the simplest example where both such saddles can appear:
\be 
Z_{1}(A,B) = \sum_{i,j} \s{p_{i}p_{j}}\; \la \psi_{i} | \psi_{j} \ra_{A}\;\la \psi_{j} | \psi_{i} \ra_{B} \big|_{{\rm gravity}}. \label{eq:normalization}
\ee
 
\begin{figure}
     \centering
     \includegraphics[scale=.2]{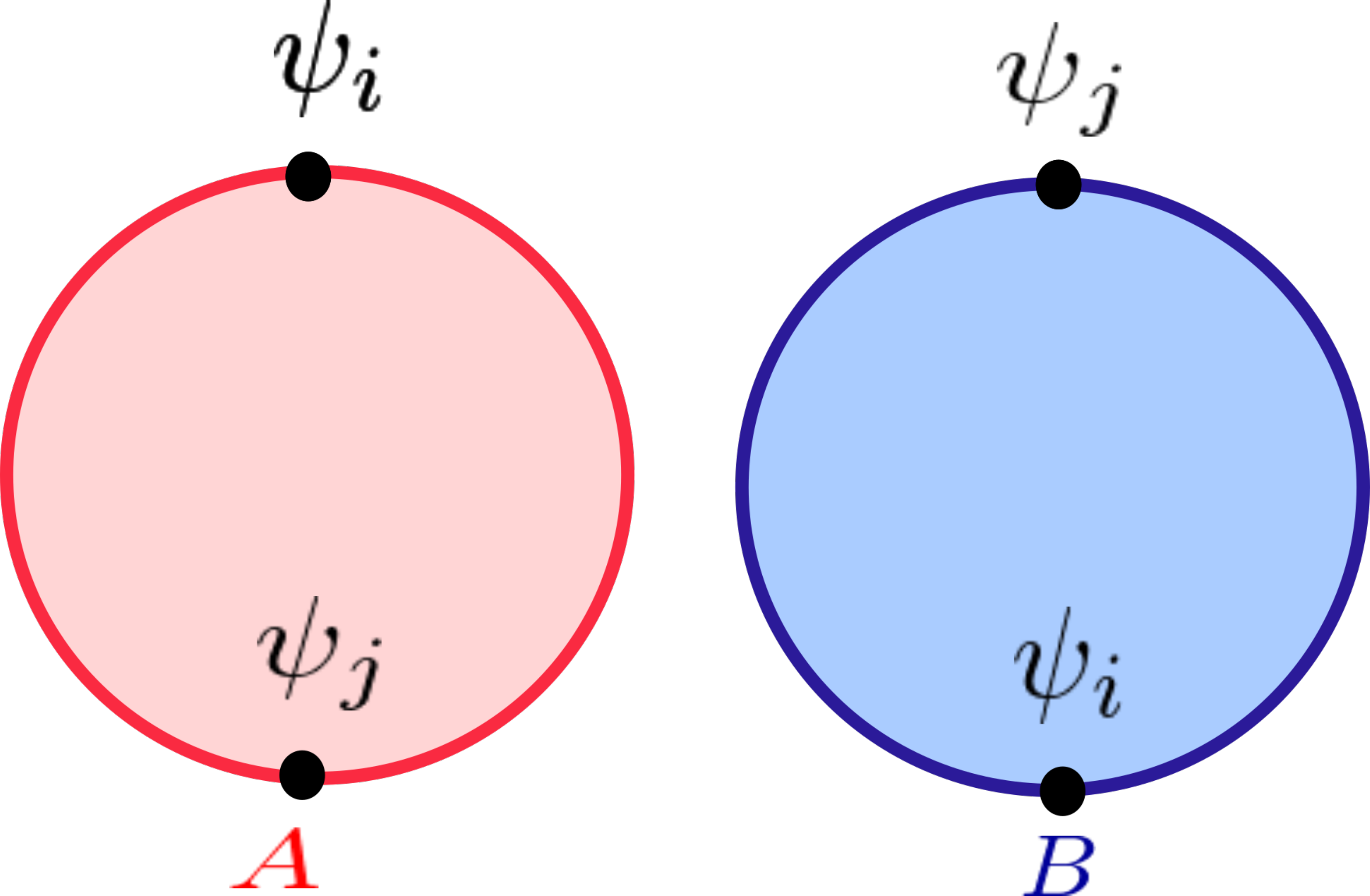}
     \hspace{3cm}\includegraphics[scale=.2]{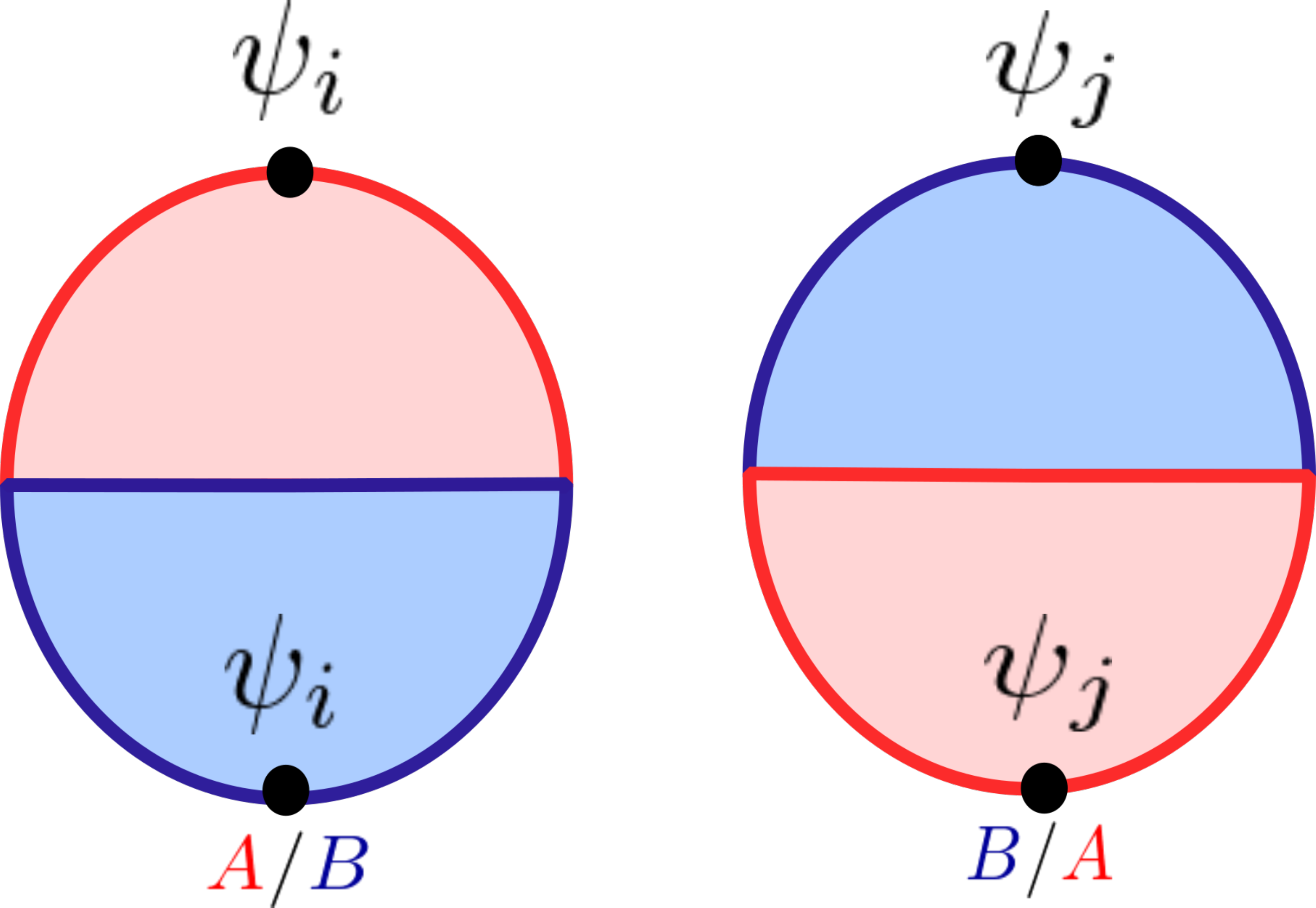}
     \caption{\small{The calculation of $Z_{1}(A,B)$ \eqref{eq:thesimplest} with the inclusion of a swap wormhole. Left: Trivial saddle $M_{1}$. Right: Saddle with a swap wormhole $M_{2}$.
     }}
     \label{fig:swap2}
     \end{figure}
This path integral appears as a normalization factor in the R\'enyi entropy \eqref{eq:renyithiscase}.
In order to emphasize that we are evaluating the product of overlaps $\la \psi_{i}| \psi_{j} \ra\la \psi_{j}| \psi_{i} \ra $ in the presence of gravity, we introduced the notation $``\big|_{{\rm gravity}} "$. 
This gravitational path integral includes contributions from at least two interesting saddle points.
The first is the trivial saddle $M_{1}$ which consists of the original disjoint universes $M_{1}: \{A,B\}$, and the second is the swap wormhole $M_{2}: \{A/B,B/A\}$ (Fig. \ref{fig:swap2}) which, as we have explained, appears as the effective description of a cylinder wormhole connecting universes $A$ and $B$.
Summing the contributions of these two saddles, we find
\begin{align}
Z_{1}(A,B)&= e^{-S_{{\rm grav}}[M_{1}]} \;\sum_{i,j} \s{p_{i}p_{j}}\;\la \psi_{i}| \psi_{j}\ra_{A} \la \psi_{j}|\psi_{i} \ra_{B}\big|_{{\rm CFT}} \nonumber \\
& \quad + e^{-S_{{\rm grav}}[M_{2}]} \;\sum_{i,j} \s{p_{i}p_{j}}\; \la \psi_{i}| \psi_{i}\ra_{A/B} \la \psi_{j}|\psi_{j} \ra_{B/A}\big|_{{\rm CFT}} .
\label{eq:thesimplest}
\end{align}
Here we introduced the notation $``\big|_{{\rm CFT}}"$ in order to emphasize that we are evaluating these overlaps in the CFT on each saddle.
These sums can be written in terms of partition functions of the CFT on a fixed hyperbolic metric, which in JT gravity is fixed. 
Notice that the ordering of the operators is a bit different than in the replica manifold $M_{2n}$ with $n>1$.
Namely, on a given cylinder sheet of $M_{2n}$, there are two identical operators $\psi_{j_k}$ and two different operators $\psi_{i_k}$ and $\psi_{i_{k+1}}$, but here there is only one sheet with two pairs of identical operators $\psi_i$ and $\psi_j$.
Therefore, we can immediately write down thermal partition function interpretations for the CFT contributions.
With this understanding, we obtain
\be 
\sum_{i,j} \s{p_{i}p_{j}}\;\la \psi_{i}| \psi_{j}\ra_{A} \la \psi_{j}|\psi_{i} \ra_{B}\big|_{{\rm CFT}}=1, \quad  
\sum_{i,j} \s{p_{i}p_{j}}\; \la \psi_{i}| \psi_{i}\ra_{A/B} \la \psi_{j}|\psi_{j} \ra_{B/A}\big|_{{\rm CFT}} =\left(\f{Z_{{\rm CFT}} \left(\f{\beta}{2} \right)}{Z_{{\rm CFT}}^{\f{1}{2}}(\beta)}\right)^{2}.
\label{eq:cftpart}
\ee
In the high temperature limit $\beta \rightarrow 0$, the CFT partition function behaves like 
\be 
Z_{{\rm CFT}}(\beta) \sim e^{\f{\pi^{2}c}{3\beta}}, \qquad \f{Z_{{\rm CFT}} \left(\f{\beta}{2} \right)}{Z_{{\rm CFT}}^{\f{1}{2}}(\beta)}  \sim e^{\f{\pi^{2}c}{2\beta}} ,
\ee
so the swap wormhole contribution dominates the disconnected contribution in \eqref{eq:cftpart} as $\beta \to 0$.
The suppression in the disconnected saddle comes from the fact that the bra $\la \psi_{i}|$ and ket $ |\psi_{j} \ra$ on each $A$ and $B$ are different, and therefore the CFT overlap is large only when  $i=j$. 
This reduces the effective number of index sums which contribute.
On the other hand, in the swap wormhole, the operator insertions are always identical, so the number of sums is not reduced. 
By writing 
\be
e^{-S_{{\rm grav}}[M_{1}]}= Z_{{\rm grav}} (A) Z_{{\rm grav}} (B), \qquad e^{-S_{{\rm grav}}[M_{2}]}=Z^{2}_{{\rm grav}} (A/B) ,
\ee
the net result for  $Z_{1} (A,B)$ is then given by 
\begin{align} 
Z_{1} (A,B) 
= Z_{{\rm grav}} (A) Z_{{\rm grav}} (B) +\left(\f{Z_{{\rm CFT}} \left(\f{\beta}{2} \right)}{Z_{{\rm CFT}}^{\f{1}{2}}(\beta)}\right)^{2} Z^{2}_{{\rm grav}} (A/B).
\end{align}

Assuming that the gravitational actions $S_{{\rm grav}}[M_{1}]$ $S_{{\rm grav}}[M_{2}]$ are bounded in the high temperature limit  $\beta \rightarrow 0$, the total gravitational path integral $Z_{1} (A,B)$ is dominated by the contribution of the swap wormhole, 
\be 
Z_{1} (A,B) =\left(\f{Z_{{\rm CFT}} \left(\f{\beta}{2} \right)}{Z_{{\rm CFT}}^{\f{1}{2}}(\beta)}\right)^{2} Z^{2}_{{\rm grav}}   (A/B), \quad \beta \rightarrow 0.
\ee
To summarize, we saw that the swap wormhole dominates the gravitational path integral \eqref{eq:thesimplest} over the disconnected contribution.
We argued previously that other wormhole configurations do not contribute more than the cylinder wormhole of Sec.~\ref{sec:cylinder-quotient}, of which the swap wormhole is an effective description.
Mechanically, the cylinder and/or swap wormhole becomes the dominant saddle because  it allows us to swaps the ket state in universe $A$ for the ket state in universe $B$, and the bra and ket states in the new disks $\{A/B, B/A \}$ are then identical. 
The saddlepoint value of the partition function is maximized when this condition is satisfied. 

\section{Entanglement entropy between two gravitating universes}
 \label{sec:twogravitating}

In Sec.~\ref{sec:cylinder-quotient} we described the zoo of wormholes that could potentially  contribute to the R\'{e}nyi entropy, and used an index contraction argument to demonstrate that the dominant piece somes from the maximally connected wormhole  $M_{2n}$. We then took a quotient by the replica $\mathbb{Z}_n$ symmetry in order to analytically continue in $n$, which led to a cylinder wormhole between the two universes  (see Fig.~\ref{fig:degenerationlimit}).   Finally in Sec.~\ref{sec:effective} we gave an  effective description of this cylinder as a swap wormhole, i.e. a disk with half of the its boundary from each universe (Fig.~\ref{fig:state-operator}).  This effective description, which is valid at high entanglement temperature is particularly convenient for the explicit computation of the entropy, as we illustrate below.

\subsection*{Evaluation of R\'enyi entropy as $n\to 1$}
Explicitly constructing such a replica wormhole saddle is in general a difficult task, but if we are only interested in evaluating the R\'enyi entropy in the $n \rightarrow 1$ limit, there is an effective prescription. 
According to the CFT state-operator correspondence, excited states are realized by insertions of local operators on the poles of the disk, so on this saddle point, the product of overlaps in the R\'enyi entropy \eqref{eq:renyithiscase} is given by combining the CFT part \eqref{eq:corrfactrized}  and the gravitational part,
\begin{align}
\prod^{n}_{k=1}
 \la \psi_{i_{k}}|\psi_{j_{k}}\ra_{A_{k}}\;  \la \psi_{j_{k}}|\psi_{i_{k+1}}\ra_{B_{k}} &=
 Z_{{\rm grav}}(B/A)^{n} \left(\prod_{k=1}^{n}  \la \psi_{j_{k}}|\psi_{j_{k}}\ra_{(B/A)_{k}} \big|_{{\rm CFT}}\right)  \nonumber \\
&\times Z_{{\rm grav}}^{(n)}(A/B)\; \left\langle \prod_{k=1}^{n} \psi_{i_{k}} (\infty_{k})  \psi_{i_{k+1}} (0_{k})  \right\rangle_{\Sigma_{n} (C)}
, \label{eq:theproduct}
\end{align}
where $Z_{{\rm grav}}(B/A)$ is the gravitational action of  $B/A$ (see Sec.~\ref{sec:swapwormhole} for a definition of the $B/A$ notation), and $Z_{{\rm grav}}^{(n)}(A/B)$ the action of $n$ copies of $A/B$  with the gluing along the cut $C$. 
The notation $\la \cdots \ra_{\Sigma_{n} (C)}$ denotes the correlation function in the CFT evaluated on the branched disk
$\Sigma_{n} (C)$. 

Let us elaborate on the above expression.  
In \cite{Balasubramanian:2020coy}, it was observed that the correlation function in \eqref{eq:theproduct} has a nice CFT R\'enyi entropy interpretation once we use the identity 
\be
\left\langle \prod_{k=1}^{n} \psi_{i_{k}} (\infty_{k})  \psi_{i_{k+1}} (0_{k})  \right\rangle_{\Sigma_{n} (C)}
=\left\langle \prod_{k=1}^{n} \psi_{i_{k}} (\infty_{k})  \psi_{i_{k}} (0_{k})  \right\rangle_{\Sigma_{n} (\overline{C})},
\label{eq:identitycorrelator}
\ee
where $\Sigma_{n} (\overline{C})$ is the branched disk with the cut on $\overline{C}$, i.e. the complement of $C$ on the time slice of $A/B$ (the hyperbolic disk). 
Intuitively, this equality holds because we can pull the cut $C$ to $\overline{C}$ while keeping its endpoints fixed, and during this process it crosses the operator on the south pole.
This has the overall effect of sending every operator at the south pole of some sheet $k$ to the south pole of sheet $k-1$.

From the right hand side of the identity \eqref{eq:identitycorrelator}, we see that when the size of the cut $|C|$ is comparable to the size of the reflection symmetric slice of the hyperbolic disk, or equivalently the size of its complement is small. $\overline{C} \rightarrow 0$.  So we have the following factorization of the correlator:
\be
\left\langle \prod_{k=1}^{n} \psi_{i_{k}} (\infty_{k})  \psi_{i_{k+1}} (0_{k})  \right\rangle_{\Sigma_{n} (C)} \rightarrow 
\la \psi_{i_{1}} | \psi_{i_{1}} \ra_{(A/B)_{1}} \la \psi_{i_{2}} | \psi_{i_{2}} \ra_{(A/B)_{2}}  \cdots   \la \psi_{i_{n}} | \psi_{i_{n}} \ra_{(A/B)_{n}}. 
\ee
Therefore, in this limit, the bra and ket states of each universe are always identical, so in the high temperature limit $\beta \rightarrow 0$,  this saddle point becomes the dominant one.
By using  the relation \eqref{eq:identitycorrelator} in the correlation function \eqref{eq:theproduct} and performing the sums over indices, the contribution of this saddle point to the gravitational path integral is given by
\be
Z_{n}(A,B)= Z(B/A)^{n} \;  Z^{(n)}(A/B),
\ee
where $Z(B/A)$ is the partition function on the manifold $B/A$, 
\be 
Z(B/A) =\left(\sum_{i} \s{p_{i}} \la \psi_{i} | \psi_{i} \ra_{B/A} \right) Z_{{\rm grav}}(B/A)= \left(\f{Z_{{\rm CFT}} \left(\f{\beta}{2} \right)}{Z_{{\rm CFT}}^{\f{1}{2}}(\beta)}\right) \;Z_{{\rm grav}}(B/A),
\ee
and $Z_{{\rm CFT}}(\beta)$ is the CFT partition function of the inverse temperature $\beta$.
The contribution of $\Sigma_n(C)$, the wormhole formed by gluing $\{(A/B)_{k}\}_{k=1}^{n}$ through the cut $C$, is given by
\begin{align}
  Z^{(n)}(A/B) &=Z^{(n)}_{{\rm grav}}(A/B) \; \sum_{\{i_{k}\}} \;   \s{p_{i_{1}} \cdots p_{i_{n}}}   \;  \la  \prod^{n}_{k=1}\psi_{i_{k}} (\infty_{k})
\psi_{i_{k}} (0_{k}) \ra_{\Sigma_{n}(\overline{C})} \nonumber \\
&= Z^{(n)}_{{\rm grav}}(A/B) \;  \f{\tr \rho_{\f{\beta}{2},\overline{C}}^{n}}{\tr \rho_{{\rm vac},\overline{C}}^{n}} \;  \left(\f{Z_{{\rm CFT}} \left(\f{\beta}{2} \right)}{Z_{{\rm CFT}}^{\f{1}{2}}(\beta)}\right)^{n} ,
\end{align}
where  $\tr \rho_{\f{\beta}{2},\overline{C}}^{n}$ and $\tr \rho_{{\rm vac},\overline{C}}^{n}$ are thermal and  vacuum R\'enyi entropy on  the interval $\overline{C}$ of the CFT respectively.  
In a CFT with large central charge limit and sparse spectrum, these are
\be 
 \tr \rho^n_{\f{\beta}{2},\overline{C_x}}=\left( \sinh \f{\pi L  (1-x)}{\beta}\right)^{-\Delta_{n}}, \quad \tr \rho^n_{{\rm vac},\;\overline{C_x}} =\left( \sin  \f{\pi^{2} x}{L} \right)^{-\Delta_{n}} ,   \quad \Delta_{n} =\f{c}{12} \left(n-\f{1}{n} \right) . \label{eq:renyinCFT}
\ee
However, as we will see later, the details of these expression (and indeed the particular CFT itself) are unimportant in the high temperature limit $\beta \rightarrow 0$ because the size of the complement $\overline{C}$ becomes small in this limit, $|\overline{C}| \rightarrow 0$.
The only relevant fact about the thermal partition function is essentially that it increases with $1/\beta$ without bound.
Combining these in the high temperature limit $\beta \rightarrow 0$ where $M_{2n}$ dominates, the expression for the R\'enyi entropy reads
\be 
\tr \rho_{A}^{n}\big|_{{\rm swap}}=\f{Z_{n}(A,B)}{Z^{n}_{1} (A,B)}=Z^{(n)}_{{\rm grav}}(A/B) \;  \f{\tr \rho_{\f{\beta}{2}}^{n}}{\tr \rho_{{\rm vac}}^{n}}.
\ee

Finally, let us take the $n \rightarrow 1$ limit of the R\'enyi entropy. 
As in the case of one non-gravitating universe and one gravitating universe, the gravitational action on the branched disk $\Sigma_{n}$ becomes proportional to the dilaton 
profile on the $n=1$ geometry,  $I_{{\rm grav}} = (n-1) \Phi_{n=1}$ for  $n \rightarrow 1$.  
Since in this situation the $n=1$ geometry is the swap wormhole $A/B$, the gravitational action is proportional to the dilaton profile on this geometry, $\Phi_{n=1} =\Phi_{A/B}$. 
Combining these, we get
\begin{align}
S_{{\rm swap}} 
=\underset{\overline{C}}{\text{min ext}}\left[ \Phi_{A/B} (\p \overline{C}) + S_{\beta/2} (\overline{C}) -S_{{\rm vac}} (\overline{C})\right] 
\label{eq:islandg}
\end{align}
where $\Phi_{A/B}$ is the dilaton profile made by gluing the boundary conditions of the two universes.
We have placed the subscript ``swap'' on this entropy to make it clear that it comes from a quantum extremal surface that lives in the swap wormhole saddle.
In the second line of \eqref{eq:islandg}, we used the purity of the state \eqref{eq:tfdstate} on $AB$.

\section{Lorentzian entropy calculation}
\label{sec:lorentzian}

So far, we have been calculating partition functions in Euclidean signature.  
Now, we would like to apply our results to Lorentzian geometries by analytically continuing the spacetime, and finding the relevant quantum extremal surface whose generalized entropy yields the entanglement entropy $-\tr \rho_A \log \rho_A$ of \eqref{eq:tfdstate} at high entanglement temperatures $\beta \to 0$.

\subsection{Analytic continuation of swap wormholes}

Since the swap wormhole is a disk, its boundary is a circle. 
We impose the boundary conditions of universe $A$ on the upper half of the circle, and those of universe $B$ on the lower half. 
Normally, when analytically continuing the Euclidean disk, we first cut the disk at its equator, then glue this section to the reflection symmetric slice of the Lorentzian geometry. 
However, the equator of the swap wormhole is a subtle concept, in the following sense. 
Let $\lambda_{A}(x)$ be the set of functions collectively denoting the boundary conditions for universe $A$, and $\lambda_{B}(x)$ be the analogous quantity for universe $B$.
Let the coordinate of the boundary circle be $x$, where $0 \leq x \leq 2\pi$. 
Then the boundary conditions  of the swap wormhole near $x=0$ are given by $\lambda_{{\rm Swap}}(x) =  \lambda_{A}(x) \theta (x) +   \lambda_{B}(x)\theta (-x)$, where $\theta (x)$ is a step function. 
In particular, this new boundary condition for the swap wormhole has discontinuities at $x=0$ and $x=\pi$, where the equator intersects with the boundary.
At one of these discontinuities, the boundary condition $\lambda_{{\rm Swap}}(x)$ is ambiguous.
At the same time, these two points are directly connected to the left and the right boundaries of the Lorentzian spacetime, so if we were to join the Euclidean and Lorentzian sections here then the Lorentzian boundary conditions would suffer from the same ambiguity.

\vspace{0.2cm}

\begin{figure}
     \centering
     \includegraphics[width=8cm]{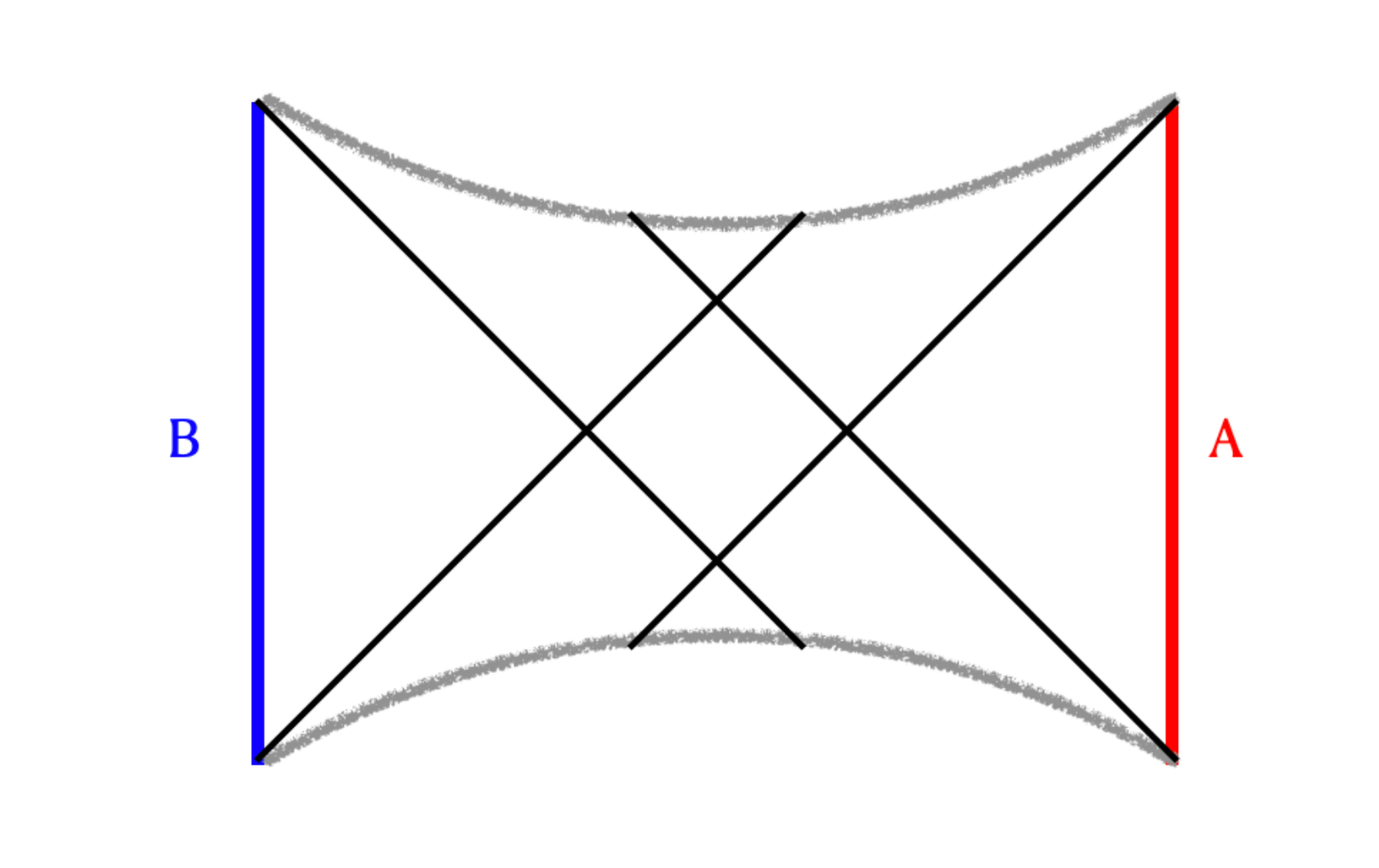}
     \caption{\small{ Lorentzian   continuation of the swap wormhole.   It is a two sided black hole with the boundary conditions of universe $A$ on one asymptotic boundary, and the conditions for the universe $B$ on the other.  We draw this Penrose diagram  from the dilaton profile \eqref{eq:dilexample}.
     }}
     \label{fig:Penrose}
 \end{figure}

\begin{figure}
     \centering
     \includegraphics[width=10cm]{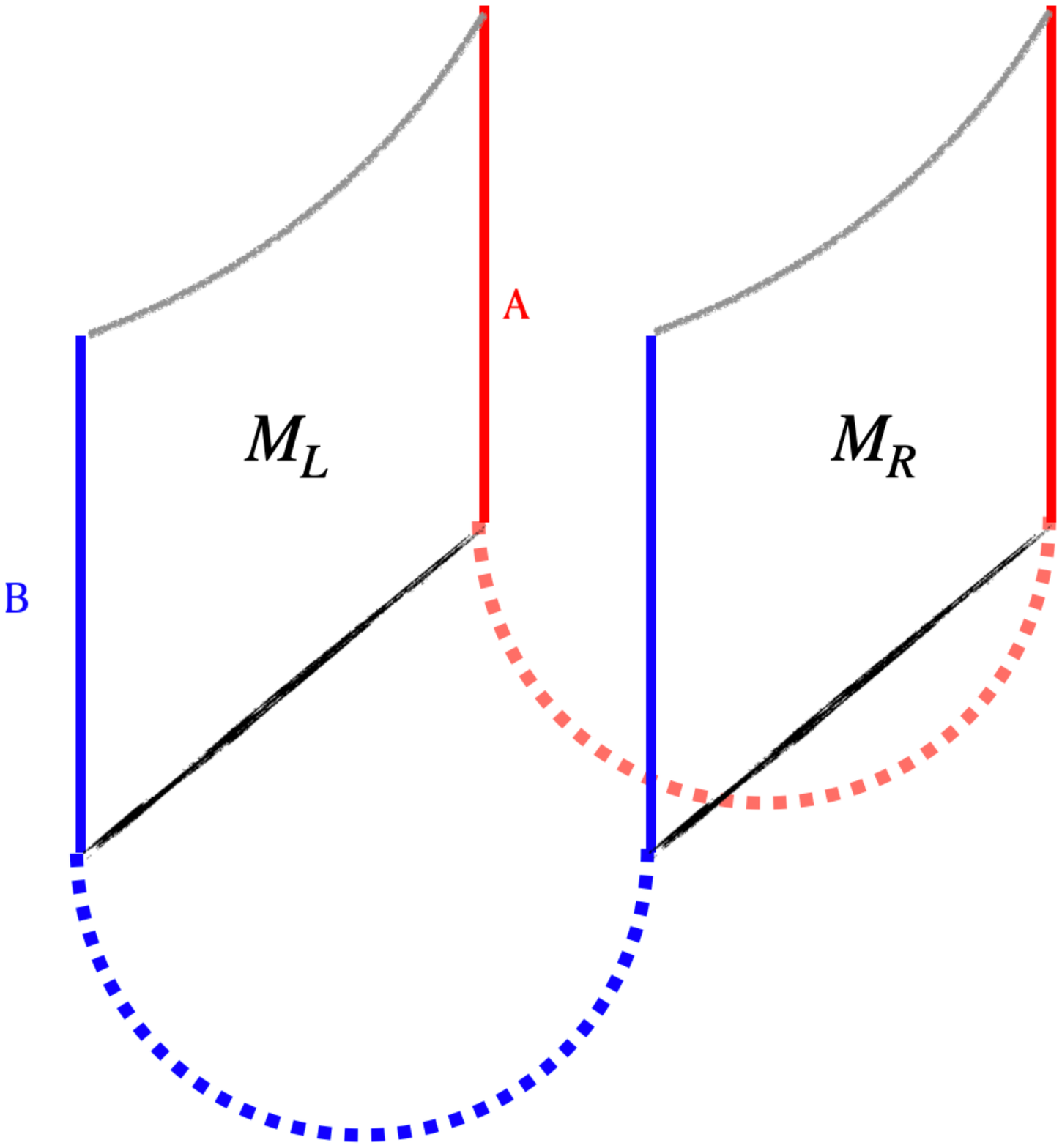}
     \caption{\small{ Analytic continuation of the cylinder wormhole. The Euclidean part (the dotted lines) is continued to the Lorentzian part which can be interpreted as two entangled black holes on the black lines.   In the short wormhole limit $\ell \rightarrow 0$, the entanglement between  them is small, so  they can be regarded as two decoupled 
black holes. Each of these black holes can be regarded as an analytic continuation of the swap wormhole.
     }}
     \label{fig:analyticcont}
 \end{figure}

A more reasonable prescription for analytic continuation is  the following. 
Instead of directly analytically continuing on  the equator, let us consider a slightly offset equator which intersects the boundary circle at $x=\varepsilon$ and $x=\pi +\varepsilon$. At these points, the boundary conditions are well defined.
Specifically, the boundary condition at $x=\varepsilon$ is simply $\lambda_{A} (x)$ and the boundary condition at $x= \pi +\varepsilon$ is $\lambda_B(x)$.
We can then consider analytic continuation from this offset equator in the limit $\varepsilon \rightarrow 0$, which will almost restore the reflection symmetry.
The resulting Lorentzian geometry is a new eternal black hole which is generated by gluing the two original eternal black holes in universes $A$ and $B$. 
More precisely, this new eternal black hole satisfies the gravitational equations of motion with boundary conditions $\lambda_{\text{Swap}}(x)$.
The Penrose diagram of the resulting geometry is shown in Fig.~\ref{fig:Penrose}.
While this offset equator slice may seem slightly ad hoc, our motivation for defining it comes from the cylinder wormhole, where the reflection symmetric slice actually does connect the boundaries of universes $A$ and $B$ without any ambiguity (Fig.~\ref{fig:analyticcont}).
As we will see below the offset continuation prescription for the swap wormhole arises directly from its interpretation as the effective description of the full cylinder wormhole.

\subsection{Analytic continuation of cylinder wormholes}  

We now show that the same Lorentzian description emerges instead from  analytically continuing the cylinder wormhole.
We first cut the cylinder $S^{1} \times I$ into two pieces by splitting the circle direction into two intervals $S^{1} = I_{N} \cup I_{N}$,   with $I_{N} :  0 \leq x \leq \pi$ and $I_{S} :  \pi \leq x \leq 2\pi$.
Then the Euclidean  gravitational path integral on the bottom piece  $I_{S} \times I$ prepares an entangled state on the two disjoint lines at $x=0$ and $x=\pi$.  
These lines are smoothly continued to the Lorentzian part of the geometry as in Fig.~\ref{fig:analyticcont}. Thus, the Lorentzian geometry consists of two disjoint pieces, $M_{R}$ at  $x=0$ and $M_{L}$ at $x= \pi$.
Each Lorentzian piece has two spatial  boundaries. 
For example, let us write $\p M_{L} =b_{1L}\cup b_{2R}$. Then we impose boundary conditions for universe $A$ on $b_{1L}$ and 
those of universe $B$ at  $b_{2L}$.
Similar conditions are imposed on the spatial boundaries of the right Lorentzian piece $M_{R}$. 

\vspace{0.2cm}

The fact that these two disjoint Lorentzian spacetimes $M_{L}$ and $M_{R}$ are connected in the Euclidean regime means that $M_{L}$ and $M_{R}$ are entangled. 
This can be clearly seen in the annulus picture of the cylinder wormhole. 
The reflection symmetric slice of the annulus has two disjoint pieces, say $T_{L}$ and $T_{R}$.   
$T_{L}$ is the time slice of $M_{L}$ and $T_{R}$ is a similar slice of $M_{R}$. 
The state on the two disjoint time slices is prepared by the Euclidean path integral on the lower part of the annulus, so it has a thermofield double-like expression where the fields on $T_{L}$ are entangled with the those on $T_{R}$.
The amount of entanglement between quantum fields on these two Lorentzian spacetimes $M_{L}$ and $M_{R}$ is small when the circumference of the annulus is large.  

\vspace{0.2cm}

Viewed another way, in the short length limit of the cylinder wormhole $\ell \rightarrow 0$, the width of the annulus becomes small. 
This implies that, if we fix the size of the spatial direction, the circumference of the annulus will become large $b \rightarrow \infty$. 
In this limit, the entanglement between $M_{L}$ and $M_{R}$ is vanishing, and they will behave like two decoupled spacetimes. 
In this limit, $M_{L}$ and $M_{R}$ are the same two sided eternal black hole on which we imposed the boundary conditions of universe $A$ on one side and those of $B$ on the other side (Fig.~\ref{fig:Penrose}).
This is precisely the picture we found by analytic continuation of the swap wormhole.

\subsection{An example: ER=EPR from a classical scalar field}

In a theory of gravity, entanglement structure can be realized as geometry \cite{VanRaamsdonk:2010pw}.
A particularly subtle case of this is the ER=EPR principle \cite{Maldacena:2013xja}, where early Hawking radiation from an evaporating black hole is highly entangled with the black hole interior.
In this setting, the ER=EPR principle says that this entanglement can be geometrized as a wormhole that connects the black hole interior to the far radiation region.
Much effort has gone into making this idea concrete, including the development of stringy models \cite{Jafferis:2021ywg}, large supergravity wormholes\footnote{For other Lorentzian wormhole constructions, see \cite{Maldacena:2018gjk,Maldacena:2020sxe}. } created by entanglement \cite{Balasubramanian:2020ffd}, and toy models which exhibit phase transitions driven by coupling \cite{Gao:2016bin,Maldacena:2017axo,Maldacena:2018lmt,Fallows:2020ugr} or entanglement \cite{Anderson:2021vof}.

In our setup, perhaps the most convincing demonstration of the ER=EPR principle in semiclassical gravity would involve taking the gravitational boundary conditions in universes $A$ and $B$ to be different.
That is to say, the metric itself would have different boundary conditions on the two sides of the wormhole, and this would correspond to something like two black holes at different temperatures.
Such a scenario presents some technical challenges which we comment on further in Sec.~\ref{sec:disc}.
In this work, we will simply demonstrate the ER=EPR principle for a wormhole with different boundary conditions on either end for a classical scalar field.
Before proceeding, we comment that the general lesson seems to be that the asymmetry in boundary conditions induces a profile for bulk fields that contributes to the stress tensor, and this extra contribution backreacts on the geometry, modifying the resulting entanglement entropy at $O(1/G_N)$ by changing the area term.

The total effective action of our gravitational system is
\be
\log Z = \log Z_{{\rm CFT}} -I_{\text{JT}} -I_{\chi} , 
\ee
where $I_\text{JT}$ is the action of JT gravity defined in \eqref{eq:action}, $\log Z_\text{CFT}$ is the partition function of the conformal matter, and $I_{\chi} $ is the action of a classical scalar field
\be
I_{\chi} = -\f{1}{2} \int dx^{2} \s{-g} g^{ab} \p_{a} \chi \p_{b} \chi .
\ee
In JT gravity, the metric is always fixed to that of pure AdS$_{2}$,
\be
ds^{2} = \f{-d\tau^{2} +d \mu^2}{\cos^{2} \mu}.
\label{eq:ads2-metric}
\ee
The AdS boundaries correspond to $\mu= \pm \f{\pi}{2}$.
We choose the gravitational configuration of universe $A$ to be an AdS eternal black hole with dilaton and scalar field profiles
\be
\Phi_{A}(\tau, \mu) = \phi_{0} +\bar{\phi} L\; \f{\cos \tau}{\cos \mu}, 
\quad \chi_{A}(\tau, \mu) = \lambda_{A} .
\ee
The coefficient $L$ appearing in the dilaton profile determines the temperature of the black hole $T$ via $T =L/2\pi$. 
Because the scalar field profile is constant, its derivative vanishes, and therefore it makes no contribution to the stress tensor.
The black hole has one bifurcation surface at $(\tau, \mu) =(0,0)$, and the dilaton value at this point corresponds to the entropy of the black hole $S_{{\rm BH}}$:
\be
S_{{\rm BH}} =\Phi(0,0) =\phi_{0} +\bar{\phi} L.
\ee
On universe $B$ we place the same eternal black hole but consider a different constant scalar field profile:
\be
\Phi_{B}(\tau, \mu) = \phi_{0} +\bar{\phi} L\; \f{\cos \tau}{\cos \mu},
\quad \chi_{B} (\tau, \mu)= \lambda_{B} .
\ee
Now we construct the new gravitational configuration on $A/B$ by gluing these profiles, as in \cite{Bak:2018txn}. A similar gluing was studied also in \cite{Bernamonti:2018vmw,Ugajin:2020dyd}.
The metric is still given by \eqref{eq:ads2-metric}.
The $A$ boundary is located at $\mu = \frac{\pi}{2}$ and the $B$ boundary is at $\mu = - \frac{\pi}{2}$.

The scalar field profile which satisfies the equation of motion $\Box \chi = 0$ and smoothly interpolates between $\lambda_{A}$ and  $\lambda_{B}$ on the boundaries is
\be 
\chi_{A/B}(\tau, \mu) = \gamma \left( \mu -\f{\pi}{2}\right) + \lambda_{A},  \quad   \gamma = \f{\lambda_{A} -\lambda_{B}}{\pi} .
\ee
To write down the equation of motion for the dilaton $\Phi$, it is convenient to use lightcone coordinates  $x^{\pm} =\tau \pm \mu$
where the equations that follow from varying the metric are
\cite{Almheiri:2014cka}
\begin{equation} 
\begin{split}
2\p_{+} \p_{-} \Phi +e^{2\omega} \; \Phi &=4\pi  \;\la T_{+-}\ra,  \\
 e^{2\omega} \p_{+} \left[  e^{-2\omega} \p_{+} \Phi \right] &=-2\pi  \; \la T_{++} \ra,  \\
  e^{2\omega} \p_{-} \left[  e^{-2\omega}  \p_{-} \Phi \right] &=-2\pi \; \la  T_{--} \ra ,
\end{split}
\label{eq:dilatoneom}
\end{equation}
with $e^{-2\omega} =\cos^{2} \mu$.  

The stress tensor in the right hand side of \eqref{eq:dilatoneom} splits into two pieces 
\be
\la T_{ab} \ra =\la T_{ab} \ra_{{\rm CFT}} +T^{\chi}_{ab},
\ee
where $\la T_{ab} \ra_{{\rm CFT}}$ is the quantum stress tensor of the CFT, and $T^{\chi}_{ab}$ is the  stress tensor of 
the classical scalar $\chi$, 
\be
T^{\chi}_{ab} = \p_{a} \chi \; \p_{b} \chi  - \f{1}{2} g_{ab} \p^{a} \chi \p_{a} \chi.
\ee

Thus in the presence of the thermal CFT matter at entanglement temperature $\beta$ and the scalar field configuration $\chi_{A/B}$, the stress-energy tensor expectation value is
\be
\la T_{\pm \pm} \ra  =\f{c}{24} \left( \f{2\pi}{\beta}\right)^{2} + \frac{\gamma^{2}}{4}, \quad \la T_{+-} \ra = -\f{c}{48 \pi} \left(\f{1}{\cos^{2} \mu} \right)
\ee
The resulting dilaton profile is 
\be
\Phi_{A/B}(\tau, \mu) =\Phi_{0}(\tau, \mu) -\f{K_{\beta,\gamma}}{2} \left( \mu \tan \mu +1\right) - \frac{c}{12},
\ee
where the constant $K_{\beta,\gamma}$ is defined as
\begin{equation}
    K_{\beta,\gamma} \equiv \frac{2\pi^2 c}{3\beta^2} +  \gamma^2 .
\end{equation}
$\Phi_{0}(\tau, \mu)$ satisfies \eqref{eq:dilatoneom} with vanishing stress energy tensor $\avg{ T_{\pm \pm} } =0$, and is given by
\be
\Phi_{0} = \alpha_{0} \f{\cos \tau }{\cos \mu}.
\ee

As in \cite{Balasubramanian:2020coy,Balasubramanian:2020xqf}, the coefficient $\alpha_0$ is determined by requiring that the asymptotic form of the dilaton near the boundaries $\mu \rightarrow \pm \f{\pi}{2}$ should match that of the original eternal black hole dilaton profile with temperature $T=L/2\pi$.
Further details on this point can be found in \cite{Balasubramanian:2020coy,Balasubramanian:2020xqf}.
In describing the full solution, it is convenient to introduce a new parameter $b$ which is related to $K_{\beta,\gamma}$ by
\be
\f{\pi}{4}K_{\beta,\gamma} = \f{\bar{\phi} L}{2} \left(b-\f{1}{b} \right) .
\ee
We caution that this parameter $b$ is unrelated to the modulus of the Euclidean cylinder wormhole discussed in previous sections.
The full dilaton solution is given by 
\be
\Phi_{A/B}(\tau, \mu) =\phi_{0} +\f{\bar{\phi} L}{2} \left(b+\f{1}{b} \right) \f{\cos \tau}{\cos \mu}-\f{\bar{\phi} L}{\pi} \left(b-\f{1}{b} \right) (\mu \tan \mu +1) - \frac{c}{12} . \label{eq:dilexample}
\ee
This dilaton profile represents an AdS eternal black hole with a long wormhole region in its interior. 
Its Penrose diagram is shown in Fig.~\ref{fig:Penrose}. 
To map out the diagram explicitly, we can locate the bifurcation surfaces, which are critical points of the dilaton profile $\p_{\mu}\Phi_{A/B}(\tau, \mu) =\p_{\tau}\Phi_{A/B}(\tau, \mu)=0$.  
There are two such surfaces at $\mu = \mu_{L}$ and $\mu= \mu_{R}$. 
The region $\mu_{L}<\mu < \mu_{R}$ in the black hole  interior corresponds the causal shadow region, which is causally inaccessible from both boundaries.
One can easily see that in the high entanglement temperature limit $\beta \rightarrow 0$, these bifurcation surfaces approach the asymptotic boundary, i.e. $\mu_{L} \rightarrow -\f{\pi}{2}$ and $\mu_{R} \rightarrow \f{\pi}{2}$, so this region in the black hole interior expands. 

\vspace{0.2cm}

Having specified the dilaton profile $\Phi_{A/B}$, let us compute $S_{{\rm swap}}$ in  \eqref{eq:islandg}.
We take the following ansatz for  $C: -\f{\pi x}{2} <\mu <\f{\pi x}{2}$ with $0<x<1$ on the reflection symmetric slice $\tau=0$. 
By symmetry, any quantum extremal surface in the geometry actually must be of this form.
By extremizing 
\be 
S_{{\rm gen}} (x) = 2 \Phi_{A/B} \left( \tau=0 , \mu = \f{\pi x}{2} \right) + S_{\beta/2} (x) -S_{{\rm vac}} (x).
\label{eq:genent}
\ee
with respect to $x$, we obtain $S_{{\rm swap}}$. Here we denote
\be
S_{\beta/2} (x)  =\f{c}{3} \log \left[ \f{\beta}{2\pi} \sinh \f{2\pi^{2}(1-x)}{\beta}\right], \quad S_{{\rm vac}} (x) =\f{c}{3} \log \left[ 2\sin \pi x \right].
\ee
This extremization procedure was performed in detail in \cite{Balasubramanian:2020coy}, and the only slight difference here is the parameter $\gamma$, which only serves to shift the constant $K_{\beta,\gamma}$ slightly.
In the high temperature limit $\beta \rightarrow 0$,  $S_{{\rm swap}}$ coincides with the Bekenstein-Hawking entropy of the original black hole, which is independent of $\gamma$.
Thus, we have
\be
S(\rho_{A}) =
\begin{cases}
 S_{{\rm no-island}} = \f{\pi^{2}c}{3\beta}  & \beta \gg \beta_{c}\\[+10pt]
  S_{{\rm swap}}= 2S_{{\rm BH}} & \beta \ll \beta_{c}.
\end{cases}   
\ee
where $\beta_{c}$ is defined by $ S_{{\rm no-island}} (\beta_{c})= 2S_{{\rm BH}}$.
Thus, we find a result consistent with unitary evolution of our entangled state \eqref{eq:tfdstate}.

Recall that in our setup we started with two disjoint quantum mechanical systems, each of which had an effective gravitational description.
Now, we have found in our calculation of the entanglement entropy $S(\rho_A)$, the relevant bulk geometry that incorporates the quantum extremal surface is actually connected between $A$ and $B$.
We have not studied the quantum gravity partition function in our JT theory coupled to CFT matter (along the lines of \cite{Maldacena:2018lmt}), so we cannot claim to have found a phase transition between bulk saddlepoints, but we have effectively found such a transition for the purpose of computing the entropy.

\subsection{Comparison to island formula}

Having found a result for the entropy $S(\rho_A)$ which is consistent with unitary evolution of the state \eqref{eq:tfdstate} using only semiclassical gravity, we ought to compare our result to other possible choices for the gravitational path integral.
It should be fairly clear from our formulae that the fully disconnected saddle for $\tr \rho_A^n$ consisting of $2n$ disks will lead to essentially the same information paradox as the one first posed by Hawking.
This saddle corresponds to simple quantum field theory on a curved spacetime with no nontrivial topology in the replica trick.
However, it is less clear what ought to be the contribution of the replica wormholes which connect some, but not all, of the $2n$ gravitational boundary conditions appearing in $\tr \rho_A^n$.
We argued in Sec.~\ref{sec:cylinder-quotient} that the fully connected wormhole $M_{2n}$ would be dominant in the limit $\beta \to 0$, but na\"{i}vely we might have expected that the type II$_A$ or II$_B$ wormholes would be sufficient to at least restore unitarity, even if they may have predicted a larger entropy than was actually present.
This intuition comes from the derivation of the island formula, which involves wormholes of precisely types II$_A$ or II$_B$, depending on which universe is non-gravitating.

However, there is a crucial subtlety in our situation which actually leads to an information paradox even if we include wormholes of type II$_A$ and II$_B$.
This subtlety actually prevents the standard island formula itself from existing in our setup even as a bound on the microscopic entanglement entropy, as the type II$_A$ and II$_B$ wormholes do not lead to such a formula.
The subtlety lies in the normalization factor $Z_1(A,B)$ which we computed in detail in Sec.~\ref{sec:swapwormhole}.
This quantity appears in the form $Z_1^n$ in the denominator of the R\'enyi entropy \eqref{eq:renyithiscase}.
In Sec.~\ref{sec:swapwormhole}, we found that there is a transition in $Z_1$ from 1 to a quantity which is exponential in $1/\beta$.
The key point is that in all types of wormholes \textit{except} the fully connected $M_{2n}$, this transition in the normalization factor leaves us with a residual dependence on $1/\beta$ in the resulting entropy.
So, if we had chosen to drop the fully connected wormhole from the gravitational path integral, we would still have an information paradox with an entropy that grows without bound.
This growth would not be as fast as Hawking's result for the entropy, as some of the $1/\beta$ dependence would be absorbed due to the wormhole connecting some of the boundary conditions, but it would nevertheless violate unitarity.
We see from this that the inclusion of $M_{2n}$ is actually not optional in our setup in order to restore unitarity semiclassically.
We have given a prescription to evaluate its contribution and found consistency with unitarity, but its presence is required on much more general grounds.
One lesson we might draw from all of this is that ER = EPR wormholes which connect gravitational subsystems are not only sufficient to restore unitarity, but they are also necessary even on top of standard island-like contributions.

Though we have argued that there is no island-like formula bounding the entropy in our model, we can still try to compare the formula we obtained for the entropy with an incorrect application of the island formula.
The island formula was applied to the geometry in question in \cite{Balasubramanian:2020coy}, and the only new feature in the situation at hand is the classical scalar field difference $\gamma$.
From our previous discussion, we see that at least for the dilaton contribution to the entropy, the effect of adding this scalar field is to increase the effective entanglement temperature via
\begin{equation}
    \beta_{\text{eff}} = \sqrt{\frac{1}{1+\frac{3\beta^2 \gamma^2}{2\pi^2 c}}} \beta .
\label{eq:beta-effective}
\end{equation}
We see that for a fixed value of $\beta$, the entropy contribution of the cylinder wormhole is actually closer to the upper limit $S_\text{BH}$ set by unitarity than the incorrect application of the island formula.
So, while neither our cylinder wormhole nor the incorrect island formula are in tension with unitarity, the island formula would under-predict the entropy.
It would be interesting to understand this feature in more generality. 
Perhaps the discrepancy between the two results can be thought of as a ``binding entropy" that must be incorporated due to the connected Cauchy slice between universes.
We must additionally keep in mind that the bulk effective entropy term in the formula \eqref{eq:genent} that we have obtained contributes a thermal entropy at a \textit{higher} temperature than that of the bulk fields themselves.
This term is subleading in the entropy in the high temperature limit, so \eqref{eq:beta-effective} is a good approximation of the overall effect there.
It would also be interesting to understand this phenomenon in more generality.
For instance, can we construct a model (possibly with more parameters) in which the entanglement temperature and the temperature of the thermal entropy contribution are arbitrarily far from each other?

\section{Discussion}\label{sec:disc}

In this paper we studied the entanglement entropy of a thermo field double type state defined on two gravitating disjoint universes.  We  computed it using the replica trick, i.e., by calculating R\'enyi entropies $\tr \rho_{A}^{n}$ and sending $n\rightarrow 1$ . The gravitational partition function   for the   R\'enyi entropies $\tr  \rho_{A}^{n}$ contains many saddles with different topology and $2n$ boundaries. We  argued that the wormhole connecting all  $2n$ boundaries  is the dominant one, when the entanglement temperature of the TFD state is large $\beta \rightarrow 0$.   The saddle point  effective action of this wormhole  has a complicated form, and even in the von Neumann limit we can not interpret  it as a generalized entropy. However, we argued that when the entanglement is very strong, we recover an island type formula for the entanglement entropy. 
The generalized entropy that appears in the new formula is evaluated on  the new spacetime  $A/B$ which  smoothly interpolates the boundary condition for the universe $A$ 
on one end, and the boundary condition for the universe $B$.  This is a concrete example of a geometric realization of quantum entanglement, in terms of wormhole in gravitating system.

\subsection{A tidal island formula}
Perhaps the most crucial point in our analysis of the fully-connected $2n$-boundary wormhole contribution to $\tr \rho_A^n$ was that we analytically continued it by quotienting using only a $\mathbb{Z}_n$ replica symmetry.
Our first instinct might have been to consider the $2n$-boundary genus zero wormhole to instead have a $\mathbb{Z}_{2n}$ cyclic replica symmetry.
However, such a symmetry would exchange boundaries associated to universe $A$ with those of universe $B$.
With this in mind, we were directly led to a cylinder quotient geometry with one boundary from universe $A$ and the other from universe $B$, the result of a quotient by $\mathbb{Z}_n$.
Therefore, the appearance of this ER = EPR wormhole between gravitating universes was an immediate consequence of the fully-connected wormhole configuration and our choice that replica symmetry should not exchange universes of different type.
The general lesson seems to be that the ER = EPR principle in the form we have described emerges naturally from the standard analytic continuation procedure for R\'enyi entropies in gravity \cite{Lewkowycz:2013nqa}, and that entanglement in the pure state \eqref{eq:tfdstate} can drive a phase transition between two Euclidean black holes and the Euclidean cylinder similar to the one studied by explicitly coupling the two theories in \cite{Maldacena:2018lmt}, at least for the purposes of computing the entropy.

The island formula was so named because it incorporated extremal gravitating subregions in entropy calculations which were disconnected from the non-gravitating region originally considered, but here we have the possibility of an emergent geometric connection between the two gravitating universes based on very general principles of the gravitational path integral and its analytic continuation.
To continue the oceanographic analogy, the island can be connected to the mainland by a bridge which emerges from the ocean when the entanglement between gravitating subregions is strong enough.
A natural name for a formula describing such an effect would be the \textit{tidal island} formula.
A tentative definition is
\begin{equation}
    S(\rho_A) = \underset{\Sigma(AB)}{\text{min}} \left( \underset{C \subset \Sigma(AB)}{\text{min ext}} \left[ \frac{\text{Area}(\partial C)}{4G_N} + S_{\text{CFT}}(C) \right] \right) ,
\label{eq:tidal-island-formula}
\end{equation}
where $\Sigma(AB)$ is a Cauchy slice of arbitrary topology that is consistent with the asymptotic boundary conditions of universes $A$ and $B$, $C$ is a subregion of $\Sigma(AB)$, and the formula assumes a saddle-point configuration has been found for any $\Sigma(AB)$ which will be included in the minimization.

\subsection{ Entanglement between two black holes with different temperatures}

As remarked briefly in Sec.~\ref{sec:lorentzian}, an interesting future direction is to consider entanglement between two black holes $BH_{1}$ and $BH_{2}$  with different temperatures $\beta_{1 }\neq \beta_{2}$, using the approach discussed in this paper.
By unitarity, the resulting entanglement entropy $S(\rho_{A})$ has to be smaller than both of the thermal entropies of the black holes $S_{BH_{1}}$ and $S_{BH_{2}}$. 
This would be a nontrivial check of the formula we derived, and perhaps a more realistic instance of the ER = EPR principle where we would essentially be modeling an early Hawking radiation particle as a small black hole entangled with the large parent black hole.
In order to do this, we first need to construct a two-sided black hole which smoothly interpolates the boundary conditions for the two black holes.
Perhaps the geometry  discussed in 
\cite{Goel:2018ubv} is related to this. Another approach to studying entangled black holes at two different temperatures, this time connected by a bath, where the microscopic definition of the states $\ket{\psi_i}$ is more manifest is considered in \cite{Khramtsov:2021xxx}.

\subsection{A holographic realization of disjoint gravitating universes}

It would also be interesting to look at a ``holographic inception" version of our setup where the bulk CFT is itself holographic as  discussed in the scenarios of \cite{Almheiri:2019hni, Balasubramanian:2020hfs,Rozali:2019day, Chen:2020uac, Chen:2020hmv,Akal:2020twv, Geng:2020fxl,Kawabata:2021hac,Geng:2021wcq,Fallows:2021sge,Anderson:2021vof,Li:2020ceg, Akal:2020ujg,Deng:2020ent}.
The dual of a two-dimensional holographic CFT in the thermofield double state \eqref{eq:tfdstate} is an eternal black hole in AdS$_{3}$.
Our 2d CFT coupled to gravity is then holographically realized on a codimension-1 brane in the three-dimensional space.  
Now we have two such gravitating universes $A$ and $B$.
Since these two universes are entangled, one of them is placed on a codimension-1 brane in the left exterior of the 3d black hole, and the other is in the right exterior. 
The main technical statement of this paper was that when the entanglement entropy between $A$ and $B$ is large, the relevant quantum extremal surface exists on a geometry $A/B$ in which $A$ and $B$ are suitably glued.  
It would be interesting to find a brane realization of $A/B$, and perhaps make contact with a recent microscopic realization of ER=EPR \cite{Jafferis:2021ywg}.

\subsection*{Acknowledgments}
We thank Kanato Goto and Tadashi Takayanagi for useful discussions. 
VB is supported in part by the Department of Energy through grant DE-SC0013528 and grant QuantISED DE-SC0020360, as well as the Simons Foundation through the It From Qubit Collaboration (Grant
No. 38559).
AK is supported by the Simons Foundation through the It from Qubit Collaboration.
TU is supported by JSPS Grant-in-Aid for Young Scientists 19K14716.

\bibliographystyle{JHEP}
\bibliography{refs}

\end{document}